\documentclass[journal]{IEEEtran}
\ifCLASSINFOpdf
\usepackage[pdftex]{graphicx}
\graphicspath{{./fig/}}
\usepackage[noadjust]{cite} 
\DeclareGraphicsExtensions{.pdf,.jpeg,.png}
\else
\fi
\usepackage{amsmath}
\usepackage{amssymb}
\usepackage{amsfonts}
\usepackage{multirow}
\usepackage{algorithm}
\usepackage{algorithmic}
\DeclareMathOperator*{\argmax}{ \operatorname{argmax}} 
\usepackage{booktabs} \newcommand{\ra}[1]{\renewcommand{\arraystretch}{#1}}
\newcommand{\myhead}[1]{\textbf{#1}}
\begin{document}
\title{A Bayesian Based Deep Unrolling Algorithm\\ for Single-Photon Lidar Systems}

\author{Jakeoung~Koo,
        Abderrahim~Halimi,~\IEEEmembership{Senior Member,~IEEE,}
        and~Stephen~McLaughlin,~\IEEEmembership{Fellow,~IEEE}
\thanks{This work was supported by the UK Royal Academy of Engineering under the Research Fellowship Scheme (RF/201718/17128) and EPSRC Grants EP/T00097X/1,EP/S000631/1,EP/S026428/1. \textit{(Corresponding author: Abderrahim Halimi.)}}%
\thanks{The authors are with the School of Engineering and Physical Sciences, Heriot-Watt University, Edinburgh, EH14 4AS, United Kingdom (e-mail: j.koo@hw.ac.uk; a.halimi@hw.ac.uk; s.mclaughlin@hw.ac.uk).}
}


\maketitle

\begin{abstract}
Deploying 3D single-photon Lidar imaging in real world applications faces multiple challenges including imaging in high noise environments. Several algorithms have been proposed to address these issues based on statistical or learning-based frameworks. Statistical methods provide rich information about the inferred parameters but are limited by the assumed model correlation structures, while deep learning methods show state-of-the-art performance but limited inference guarantees, preventing their extended use in critical applications. This paper unrolls a statistical Bayesian algorithm into a new deep learning architecture for robust image reconstruction from single-photon Lidar data, i.e., the algorithm's iterative steps are converted into neural network layers. The resulting algorithm benefits from the advantages of both statistical and learning based frameworks, providing best estimates with improved network interpretability. Compared to existing learning-based solutions, the proposed architecture requires a reduced number of trainable parameters, is more robust to noise and mismodelling effects, and provides richer information about the estimates including uncertainty measures. Results on synthetic and real data show competitive results regarding the quality of the inference and computational complexity when compared to state-of-the-art algorithms.
\end{abstract}


%
\IEEEpeerreviewmaketitle

\section{Introduction} \label{sec:introduction}

\IEEEPARstart{S}{ingle-photon} light detection and ranging (Lidar) is an emerging technique for reconstructing and analyzing 3D scenes and has a wide range of applications~\cite{Wallace_TVT2020,rapp2020advances}. Using time correlated single-photon counting (TCSPC) technology~\cite{buller2007ranging}, a single-photon Lidar system builds a histogram of photon counts with respect to their time-of-flights (ToF). Detecting reflected photons relies on a single-photon sensitive detector known as solid-state single-photon avalanche diode (SPAD), while ToFs are obtained by measuring the time difference between the emission of laser pulses and the detection of reflected photons.  The acquired histogram contains depth and reflectivity information about the observed objects, and reconstructing such 3D information from single-photon Lidar data has been a subject of very active research~\cite{Wallace_TVT2020,rapp2020advances}.

Two reconstruction approaches of 3D scenes from single-photon Lidar data have been widely studied: a statistical approach and a data-driven approach. Statistical methods design a statistical model with some prior information and reconstruct 3D scenes, using strategies such as optimization with spatial regularization~\cite{shin2015photonefficient,halimi2016restoration,Pawlikowska_OE2017,rapp2017few,halimi2020robust,Tobin_SR21}, Markov chain Monte-Carlo (MCMC)~\cite{hernandez-marin2008multilayered,Halimi_TCI2017b,tachella2019bayesian}, expectation-maximization~\cite{altmann2018range,legros2020expectationmaximization} or Plug-and-Play methods~\cite{venkatakrishnan2013plugandplay,tachella2019realtime}. Such methods provide good interpretability in the sense that we can predict the results, depending on the considered observation model and imposed priors, but they often require user-defined parameters and hand-crafted priors.  
Data-driven approaches using deep learning  have recently gained in popularity for single-photon Lidar systems. Existing deep learning algorithms~\cite{lindell2018singlephoton,peng2020photonefficient,sun2020spadnet,ruget2021robust} train neural networks from simulated data and aim to generalize to unseen data. Lindell et al.~\cite{lindell2018singlephoton} first proposed an end-to-end deep learning model which infers depth profiles from Lidar data. Peng et al.~\cite{peng2020photonefficient} suggested a non-local network and showed a clear benefit for low photon and high noise cases. Despite their excellent performance on challenging data these deep learning methods lack interpretability, require long running times for high dimensional Lidar data and can present over-smoothing artifacts around 3D surface boundaries.

In this paper, taking advantage of statistical and deep learning approaches, we propose an interpretable and efficient deep learning architecture for high dimensional Lidar data. We design a neural network architecture by unfolding an iterative Bayesian algorithm~\cite{halimi2021robust} in the sense that we replace some of its internal operations with neural network blocks. The proposed method is in line with an emerging technique called algorithm unrolling~\cite{gregor2010learning,monga2021algorithm}. This technique replaces the steps of a conventional iterative method by neural network blocks, hence exploiting the domain knowledge when designing the network. Following the algorithm unrolling, the proposed neural network is made interpretable via the connection to the underlying Bayesian algorithm~\cite{halimi2021robust} and is efficient in terms of the number of network parameters and running time.

Using multiscale information to process single-photon data is an essential component of several state-of-the-art 3D Lidar reconstruction algorithms. This was exploited in the proposed network which requires an initial estimate of few multiscale depths as input, instead of the large histogram data cube as in~\cite{lindell2018singlephoton,peng2020photonefficient}. In this way, the knowledge of the system impulse response function (IRF) is exploited to generate the multiscale depths and the high dimensional data is compressed to the essential information. The network layers mimic the iterative steps of the Bayesian algorithm in \cite{halimi2021robust}, which alternated between a weighted median step to choose the best depth scale to represent a pixel, and a soft-thresholding step to account for spatial correlations between pixels.  Our conversion relies on a popular tool called attention~\cite{vaswani2017attention,hu2020squeezeandexcitation,wang2018nonlocal,woo2018cbam,zhao2020efficient}, which computes weights highlighting the features or areas of interest (i.e., areas requiring attention). An attention layer is said to be hard attention~\cite{xu2015show} if the attention weights are sparse (one-hot encoding), or soft otherwise. In this paper, inspired by the weighted median filter used in \cite{halimi2021robust} and promoting sharp surface's boundaries, we consider hard attention to select the best depth scale per pixel, i.e., the one showing the highest attention weight. The proposed network also includes soft attention to improve the 3D object reconstruction by considering local spatial correlations. Results on simulated and real data show the benefit of this model when compared to the state-of-the-art learning-based algorithms~\cite{lindell2018singlephoton,peng2020photonefficient}, as it preserves surface edges, has a lower computational cost (in terms of memory or computational time) and provides uncertainty maps on the predicted depth. The uncertainty maps are obtained by connection to the underlying Bayesian method~\cite{halimi2021robust} without additional complexity, while some previous works~\cite{gal2016dropout,lakshminarayanan2016simple} require multiple passes of inference and averaging steps to predict the uncertainty of the network's  outputs.

In summary, the contributions of this paper are:
\begin{itemize}
\item an efficient deep learning model suitable for high-dimensional single-photon Lidar data,
\item interpretable neural network blocks, providing uncertainty information on the final depth map,
\item a scale selection strategy through a combination of hard and soft attention, showing competitive results when compared to state-of-the-art methods, i.e.,  less artifacts on surface boundaries, and improved robustness to mismodelling effects.
\end{itemize}

The remainder of the paper is organized as follows. Section~\ref{sec:multiscale} describes the multiscale observation model for single-photon Lidar measurements. Section~\ref{sec:underlying} reviews the underlying iterative method~\cite{halimi2021robust} resulting from a Bayesian hierarchical model. In Section~\ref{sec:proposed}, we present the proposed unrolling model with details on the training procedures. In Section~\ref{sec:experimental}, we analyze the proposed network and evaluate the performance of our method on simulated data as well as real data. Section~\ref{sec:conclusion} presents the conclusions and future work. 

\section{Multiscale observation model} \label{sec:multiscale}

This section presents the considered Poisson-based observation model for single-photon Lidar system, which is required to derive the underlying Bayesian algorithm in Section~\ref{sec:underlying}. Akin to~\cite{halimi2021robust}, we include multiscale information in the observation model.
Single-photon Lidar systems provide range information about the scene by measuring the time difference between emission of light pulses and detection of photons. Collecting such time delays, the Lidar system builds a histogram of counts denoted by $y_{n,t} \in \{0, 1, 2, \cdots \}$ where $n$ represents the pixel index and $t$ the time bin index. The observed photon counts are commonly assumed to follow the Poisson distribution with mean value $s_{n,t}$ as follows $y_{n, t} \sim \mathcal{P}\left(s_{n, t}\right)$~\cite{shin2015photonefficient,rapp2017few}. Assuming one target per each pixel $n$, the observation model for $s_{n,t}$ can be written as
\begin{equation} \label{eq:snt}
s_{n, t}=r_{n} g \left(t-d_{n}\right)+b_{n},
\end{equation}
where $r_n$ is the reflectivity of the target, $d_n$ the depth information of the target, $b_{n}$ the background photons due to ambient light and  detector dark counts and $g$ is the system IRF. We approximate the system IRF $g$ by the Gaussian function $\mathcal N(t \, ; \, \mu,\sigma^2)$ with the mean $\mu$ and the standard deviation $\sigma$ and consider that $\sum_{t=1}^T g\left(t-d_{n}\right)=1$ for all $n$, for all possible depths of the scene~\cite{halimi2016restoration,halimi2021robust}.
By assuming independent observations between $y_{n,t}, \forall n,t$, the joint likelihood for $\boldsymbol{Y}=\{ y_{n,t}\}$ can be written as
\begin{equation} 
p\,(\boldsymbol{Y} \mid \boldsymbol{d}, \boldsymbol{r}, \boldsymbol{b})=\prod_{n=1}^{N} \prod_{t=1}^{T} \frac{s_{n, t}^{y_{n, t}}}{y_{n, t} !} \exp ^{-s_{n, t}},
\end{equation}
where $\boldsymbol{d}$, $\boldsymbol{r}$, $\boldsymbol{b}$ represent the column vectors of size $N$ gathering depth, reflectivity and background parameters, respectively. Without background photons, the maximum likelihood estimate of the reflectivity can be computed as $r_n^{\mathrm{ML}}  = \bar{s}_{n}=\sum_{t=1}^{T} y_{n, t}$ and the depth as
\begin{equation} 
d_n^{\mathrm{ML}} = \arg\max_d \sum_t y_{n,t} \log g(t-d).
\end{equation}

In this case, the likelihood can be written to be proportional to the following (See Appendix of \cite{halimi2021robust} for the details)
\begin{equation} \label{eq:joint2}
\begin{aligned}
p \, ( \boldsymbol y_{n} \mid r_{n}, d_{n} ) & \propto \mathcal{G}\left(r_{n} ; 1+\bar{s}_{n}, 1\right) Q \left( \boldsymbol{y}_{n}\right) \\
& \times \mathcal{N} (d_{n} ; d_{n}^{\mathrm{ML}}, \bar{\sigma}^{2}),
\end{aligned}
\end{equation}
where $\mathcal G(x \,; \, \cdot, \cdot)$ is the gamma distribution with shape and scale parameters, $Q$ is a function of $\boldsymbol{y}_n$ and $\bar \sigma^2 := \sigma^2 / \bar s_n$. 
To handle high noise in Lidar data, it is common to incorporate multiscale information, as is done in statistical methods~\cite{rapp2017few,halimi2021robust} as well as deep learning works~\cite{lindell2018singlephoton,peng2020photonefficient,ruget2021robust}. We employ a similar multiscale approach, using the fact that low-pass filtered histograms (resulting in summing neighbouring pixels) still follow a Poisson distribution. We generate $L$ downsampled histograms $\boldsymbol{y}_n^{(\ell)}$ with $\ell \in \{2,\cdots,L\}$, by spatially downsampling the original histogram data $\boldsymbol{y}_n^{(1)}:=\boldsymbol{y}_n$ with uniform filters. This multiscale data can be efficiently computed using convolution with different uniform kernel sizes. Assuming the same observation model in~\eqref{eq:joint2}, the likelihood for each downsampled histogram $\boldsymbol y_n^{(\ell)}$ can be written as
\begin{equation} \label{eq:p(y_n^l)}
\begin{aligned}
p\left( \boldsymbol y_{n}^{(\ell)} \mid r_{n}^{(\ell)}, d_{n}^{(\ell)}\right) & \propto \mathcal{G}\left(r_{n}^{(\ell)} ; 1+\bar{s}_{n}^{(\ell)}, 1\right) Q \left( \boldsymbol{y}_{n}^{(\ell)}\right) \\
&{\times}  \mathcal{N}\left(d_{n}^{(\ell)} ; d_{n}^{\mathrm{ML}(\ell)}, \bar{\sigma}^{2(\ell)}\right),
\end{aligned}
\end{equation}
where $\bar{s}_{n}^{(\ell)}=\sum_{t=1}^{T} y_{n, t}^{(\ell)}$ and $\bar{\sigma}^{2(\ell)}   = \sigma^2/ \bar{s}_{n}^{(\ell)}$. For example, we can consider $L=4$ scales with different kernel sizes such as $1{\times}1$, $3{\times}3$, $7{\times}7$ and $13{\times}13$.

\section{Underlying Bayesian algorithm} \label{sec:underlying}

In this section, we review an underlying Bayesian algorithm proposed by Halimi et al.~\cite{halimi2021robust} which inspired the design of our deep learning method in Section~\ref{sec:proposed}. This method~\cite{halimi2021robust} follows a Bayesian approach, by considering prior distributions on the unknown depth as well as their uncertainty information. The prior distributions will be combined with the observation model in~\eqref{eq:p(y_n^l)} to derive the posterior distribution, which contains rich information regarding the parameters of interest. To exploit this distribution, the method in \cite{halimi2021robust} approximated the parameter's maximum-a-posteriori (MAP) estimator using a coordinate descent method. Although this method can estimate both depth and reflectivity, for the purpose of this paper, we only consider estimating depth profiles. 

\subsection{Prior and posterior distribution}

The observation model for multiscale depths $\boldsymbol d^{(\ell)}$ is derived in \eqref{eq:p(y_n^l)}. From this multiscale information, the goal now is to estimate the true depth denoted by a latent variable $\boldsymbol x$. On this latent variable a prior is imposed, requiring spatial smoothness within a homogeneous surface while preserving the discontinuity around the boundaries of the surfaces. To satisfy this requirement and estimate a robust depth map, Halimi et al.~\cite{halimi2021robust} introduced some pre-defined weights called \textit{guidance weights} between local pixels for each scale. A high value of $w_{n',n}^{(\ell)}$ encourages the latent variable $x_n$ to be similar to $d_{n'}^{(\ell)}$.
Using the guidance weights, the latent variable $\boldsymbol x$ is assigned the conditional Laplace distribution
\begin{equation} \label{eq:xndn}
\begin{array}{c}
x_{n} \mid d_{\nu_{n}}^{(1, \cdots, L)}, w_{\nu_{n}, n}^{(1, \cdots, L)}, \epsilon_{n} \sim \\
\prod_{n^{\prime} \in \nu_{n}}\left[\prod_{\ell=1}^{L} \mathcal{L}\left(x_{n} ; d_{n^{\prime}}^{(\ell)}, \frac{\epsilon_{n}}{w_{n^{\prime}, n}^{(\ell)}}\right)\right]
\end{array}
\end{equation}
where $\mathcal L(\,\cdot\, ; \mu, \psi)$ is the Laplace distribution with the mean $\mu$ and the scale parameter $\psi$, $\nu_n$ represents the local neighbourhood around the $n$th pixel and $\epsilon_n$ is the variance of the depth $x_n$. To ensure the positivity of the variance $\boldsymbol \epsilon$, it is assigned a conjugate inverse gamma distribution as
\begin{equation}
\boldsymbol \epsilon \sim \prod_n \mathcal{I} \mathcal{G}\left(\epsilon_{n} ; \alpha_d, \beta_d \right)
\end{equation}
where $\alpha_d$ and $\beta_d$ are user set positive hyperparameters.
Combining the prior distributions in (6) and (7) and the likelihood in (5), the posterior distribution reduces to 
\begin{equation}
\begin{aligned}
p\,(\boldsymbol{x}, \boldsymbol{\epsilon}, \boldsymbol{D} \mid \boldsymbol{Y}, \boldsymbol{W}) \propto \,& p\,(\boldsymbol{Y} \mid \boldsymbol{D}) \, p\,(\boldsymbol{x}, \boldsymbol{D} \mid \boldsymbol{\epsilon}, \boldsymbol{W}) \, p \,(\boldsymbol{\epsilon})
\label{eqt:Posterior}
\end{aligned}
\end{equation}
%
where $\boldsymbol{W}$ represents the guidance weights and $\boldsymbol{D}$ represents the multiscale depths $\boldsymbol d^{(1,\cdots,L)}$. 

\subsection{Iterative algorithm}

To approximate the parameter's MAP estimates, a coordinate descent method is employed to minimize the negative log-posterior of~\eqref{eqt:Posterior}. The algorithm proposed in \cite{halimi2021robust} updates one variable at a time while fixing other variables and is summarized in Algorithm~\ref{alg1}. 
The updates of unknown variables can be divided into three parts. Firstly, the latent variable $x_n$ is updated using a weighted median filtering as follows
\begin{equation} \label{eq:xnargmin} 
x_n  \leftarrow \underset{x}{\operatorname{argmin}}  \, \mathcal{C}(x) = \sum_{l, n^{\prime} \in \nu_{n}} w_{n^{\prime}, n}^{(\ell)}\left|x-d_{n^{\prime}}^{(\ell)}\right|.
\end{equation}
%
This operation will be replaced by attention mechanisms in the proposed deep learning model in Section~\ref{sec:proposed}.
Secondly, the multiscale depths $\boldsymbol d^{(1,\cdots,L)}$ are updated by minimizing the negative log-conditional distributions of $\boldsymbol D$ in~\eqref{eqt:Posterior} as follows:
\begin{equation}
d_{n}^{(\ell)}  \leftarrow \underset{d}{\operatorname{argmin}} \frac{\left[d-d_{n}^{\mathrm{ML}(\ell)}\right]^{2}}{2 \bar{\sigma}^{2(\ell)}}+\sum_{n^{\prime} \in \nu_{n}} \frac{w_{n, n^{\prime} \in \nu_{n}}^{(\ell)}\left|d-x_{n^{\prime}}\right|}{\epsilon_{n^{\prime}}}. \label{eq:dnargmin}
\end{equation}
This operator is known as a generalized soft-thresholding operator and can be solved analytically~\cite{parikh2014proximal}. Lastly, given the estimations of $\boldsymbol x$ and $\boldsymbol d$, the depth uncertainty information can be evaluated by considering the depth variance.   The conditional distribution of $\boldsymbol \epsilon$ is given by 
\begin{equation} \label{eq:epsilon_n}
\epsilon_{n} \mid \boldsymbol{x}, \boldsymbol{D}, \boldsymbol{W} \sim \mathcal{I} \mathcal{G}\left[L   \bar{N}+\alpha_{d}, \, \mathcal{C}\left(x_{n}\right)+\beta_{d}\right],
\end{equation}
where $\bar N=|\nu_n|$ is the number of neighbors considered. The mode of this distribution  represents the MAP estimator of $\epsilon_n$ and is given by
\begin{equation} \label{eq:epsnCxn}
\hat{\epsilon}_{n} \leftarrow (\mathcal{C}\left(x_{n}\right)+\beta_{d}) / (L  \bar{N}+\alpha_{d}+1).
\end{equation}
This formula will subsequently provides a basis to estimate the uncertainty of the depth map estimated by the neural network.
As mentioned in the previous subsection, the guidance weights $\boldsymbol W$ connect the latent variable $\boldsymbol x$ to multiscale depths. These weights play an important role in the performance of the algorithm. Halimi et al.~\cite{halimi2021robust} determines the weights $\boldsymbol W$ based on the deviation of $ \boldsymbol{d}^{\mathrm{ML}(\ell)}, \forall \ell$, from a given reference depth map, while this paper proposes to learn them  from the data, as described in the following section.

\begin{algorithm}[ht]
\caption{Iterative Bayesian algorithm~\cite{halimi2021robust}} \label{alg1}
\begin{algorithmic}[1]
       \STATE \underline{Input}: Lidar data $Y$, the number of scales $L$
       \STATE Construct downsampled histograms $\boldsymbol Y^{(\ell)}$, $\ell=1,...,L$
       \STATE Compute the multiscale depths $\boldsymbol{d}^{\mathrm{ML}(\ell)}, \forall \ell$
       \STATE Compute the guidance weights $\boldsymbol W$
       \WHILE{not converge}
		\STATE Update the variable $\boldsymbol x$ by \eqref{eq:xnargmin} 
		\STATE Update the multiscale depths $\boldsymbol d^{(1,\cdots, L)}$ by \eqref{eq:dnargmin} 
		\STATE Update the uncertainty information by \eqref{eq:epsnCxn} 
               \STATE \textbf{break} if the convergence criteria are satisfied
       \ENDWHILE
   
	\STATE \underline{Output}: $\boldsymbol x, \boldsymbol \epsilon$  
\end{algorithmic}
\end{algorithm} 

\section{Proposed unrolling method} \label{sec:proposed}

\begin{figure*}[t]
\centering
\includegraphics[width=500pt]{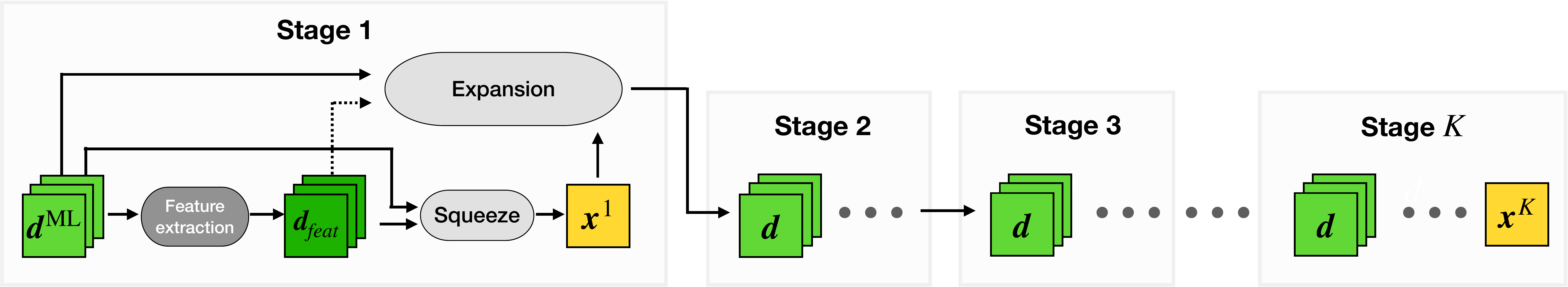}
\caption{Overview of the proposed network. The network consists of $K$ stages and each stage inputs a tuple of multiscale depths ($\boldsymbol d^{\mathrm{ML}}$), estimates a squeezed depth ($\boldsymbol x$) and refines the multiscale depths. The final stage's output of the network is a squeezed depth ($\boldsymbol x^K$). For the illustration, the network is shown for the case of three multiscales $L=3$.}
\label{fig:network}
\end{figure*}

Motivated by the Bayesian method in Algorithm~\ref{alg1}, we propose an interpretable neural network model by algorithm unrolling. As mentioned in Section~\ref{sec:introduction}, the main idea of algorithm unrolling is to unfold an underlying iterative method and mimic its operations with neural network blocks. Here, we replace the operations of Algorithm~\ref{alg1} by neural network layers. The major components of the proposed network use attention modules, which allow learning the weights $\boldsymbol W$, i.e., the correlations between local pixels at the multiscale depths.

\def\fw{200pt}
\begin{figure}[t]
\centering
\includegraphics[width=\fw]{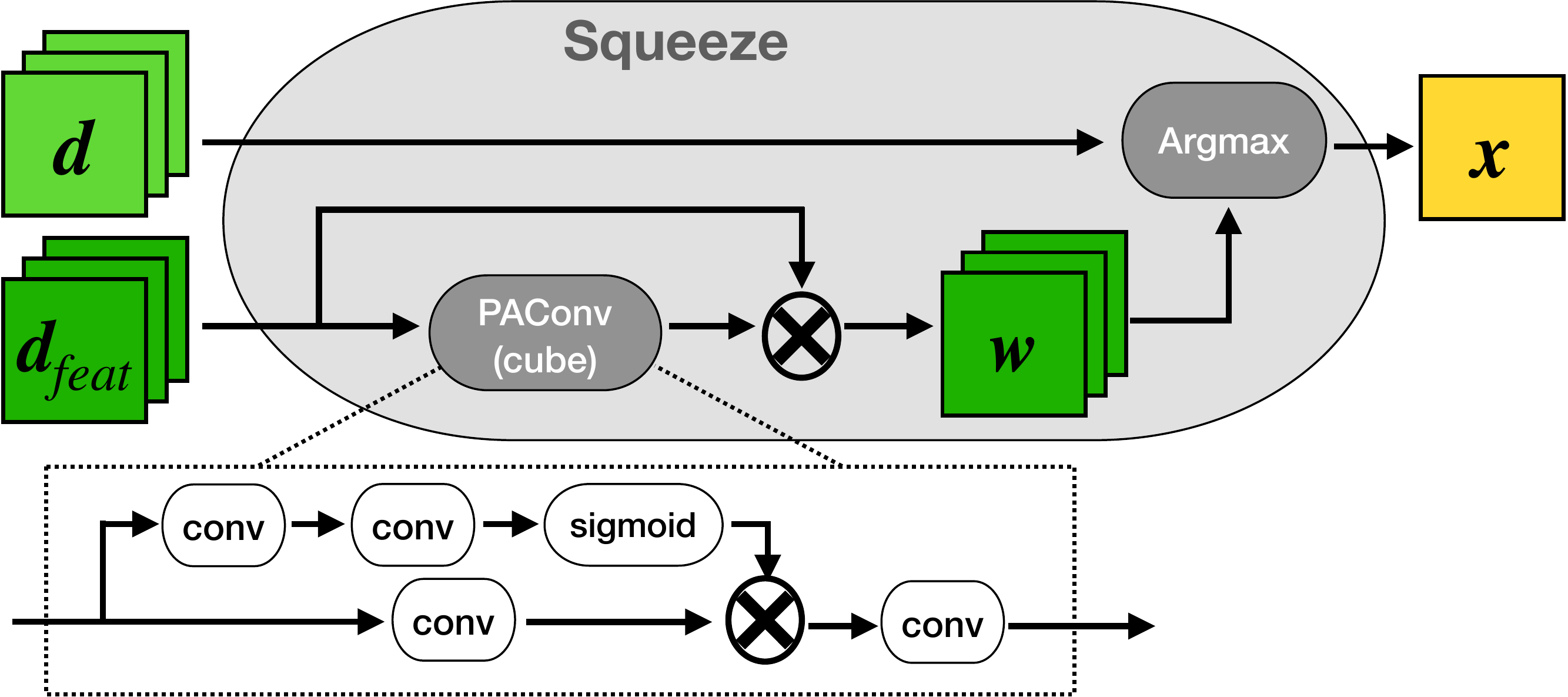}
\caption{The squeeze block estimates a squeezed depth $(\boldsymbol x)$ from a tuple of multiscale depths $(\boldsymbol d)$ and their features $(\boldsymbol d_{feat})$. The symbol $\otimes$ denotes elementwise multiplication. }
\label{fig:network2}
\end{figure}

\begin{figure}[t]
\centering
\includegraphics[width=230pt]{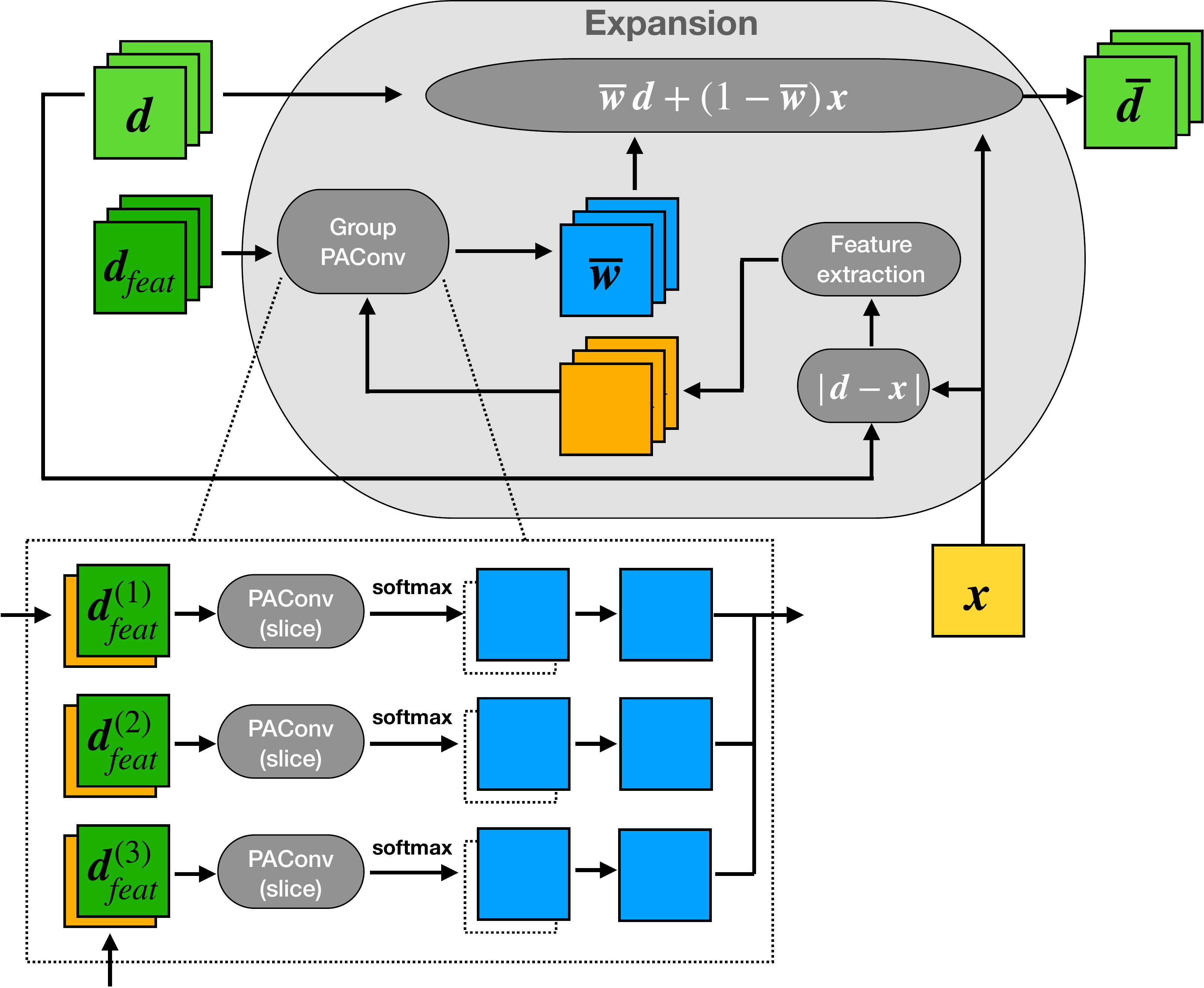}
\caption{The expansion block refines the previous multiscale depths ($\boldsymbol d$), by comparing their features ($\boldsymbol d_{feat}$) with the feature of the absolute difference between multiscale depths and the squeezed depth ($|\boldsymbol d - \boldsymbol x|$). This block outputs the refined multiscale depths ($\overline{\boldsymbol d})$.}
\label{fig:network3}
\end{figure}

\subsection{Network}

Fig.~\ref{fig:network} gives an overview of the proposed neural network. The network inputs an initial estimation of multiscale depths $\boldsymbol d^{\textrm{ML}(1,2,\cdots,L)}$ (See Section~\ref{sec:estimation}); and outputs the estimated depth $\boldsymbol x$. The network consists of $K$ stages where each stage has the same structure (except for the last stage) with different network parameters and is designed to resemble one iteration of Algorithm~\ref{alg1}. Each stage begins with the feature extraction step having three consecutive convolution layers. After that, each stage has two main blocks: \textit{squeeze} and \textit{expansion}. In this subsection, for the simplicity of notation, we use the variable symbols for the first stage and omit the dependency on the stage $k$ unless explicitly mentioned.

\subsubsection{The squeeze block }
This is a key element in the network as it estimates a single depth by using the multiscale depths and their features, as shown in Fig.~\ref{fig:network2}. This block is inspired from the weighted median filtering step~\eqref{eq:xnargmin} in the underlying Bayesian algorithm. It considers hard attention~\cite{xu2015show} to select the scale with the highest attention weight for each pixel, and takes the single depth value on that scale. Specifically, the squeeze block first computes attention weights in a module named~\textit{PAConv (cube)}, a variant of the so-called pixel attention~\cite{zhao2020efficient}. Pixel attention considers attention weights at a pixel level, so that the attention weights have the same size as the input. The module \textit{PAConv (cube)} computes  internal weights which are multiplied by the depth features, yielding the attention weights $\boldsymbol w^{(1,2,\cdots,L)}$. 
The latter weights indicate the importance of each scale, and only one scale is chosen by the argmax operation, yielding the squeezed depth. Formally, the squeezed depth for the $n$th pixel is computed by
\begin{equation} \label{eq:argmax}
x_n = d_n^{(\ell ')}, \quad \ell' = \argmax_{\ell \in \{1,\cdots,L \}} w_n^{(\ell)},
\end{equation}
where $w_n^{(\ell)}$ denotes the attention weight for the $\ell$th scale. Since the argmax operation is not differentiable, we replace it with the alternative differentiable Gumbel-SoftMax~\cite{maddison2017concrete,jang2017categorical}.

\subsubsection{The expansion block }
This block refines multiscale depths to obtain $\overline{\boldsymbol{d}}$ as the weighted average between multiscale depths ${\boldsymbol{d}} $ and the squeezed depth ${\boldsymbol{x}} $. This expansion block corresponds to the soft-thresholding step~\eqref{eq:dnargmin} in the Bayesian algorithm, as it updates the multiscale depths based on the obtained squeezed depth. However, we design the expansion block to exploit the attention framework, rather than mimicking exactly the soft-thresholding operator~\eqref{eq:dnargmin}. As shown in Fig.~\ref{fig:network3}, the expansion block has three inputs: the multiscale depths, their features $\boldsymbol d_{feat}$ and the squeezed depth $\boldsymbol x$. 
As indicated by the weighted average, the weights will help combine values from either the outlier-free squeezed depth ${\boldsymbol{x}}$ or the multiscale depths ${\boldsymbol{d}}^{(\ell)}$, where ${\boldsymbol{x}}$ values will be promoted if the two depths are significantly different. This highlights the importance of the absolute difference between the multiscale depths and the squeezed depth for each scale (i.e., $\forall \ell, \, |\boldsymbol d^{(\ell)} - \boldsymbol x|$). The latter difference is fed together with the multiscale depth features $\boldsymbol d_{feat}$ into a module named \textit{Group PAConv} to compute the weights. 
This module consists of $L$  independent sub-modules, where the $l$th sub-module inputs one scale depth feature $\boldsymbol d_{feat}^{(\ell)}$ and the difference feature   $|\boldsymbol d^{(\ell)} - \boldsymbol x|$. From such input, an attention module named \textit{PAConv (slice)} estimates internal weights for each scale. A softmax operator is then applied to these internal weights after multiplying them by a coefficient $\rho$ (note that $\rho$ is introduced to enforce weights sparsity and is fixed to $\rho=2$ throughout the paper). The softmax operator outputs two normalized channels per scale, we only consider the first channel related to the multiscale depth parts and denoted by $\overline{\boldsymbol w}$ (See the blue rectangles in the bottom part of Fig.~\ref{fig:network3}). Then, the expanded depth $\overline{d}_n^{(\ell)}$ for the $\ell$th scale and the $n$th pixel is obtained by the convex combination of $d_n^{(\ell)}$ and $x_n$ as follows:
\begin{equation} \label{eq:overline d}
\overline d^{(\ell)}_n = \overline w_n^{(\ell)} d^{(\ell)}_n + (1-\overline w_n^{(\ell)})x_n, \quad 0 \leq \overline w_n^{(\ell)} \leq 1.
\end{equation}

We have described one stage of the network corresponding to one iteration in Algorithm~\ref{alg1}. All stages have the same structure, except the last which only has the squeeze block to produce the final output of the network.

\subsubsection{Network learnable parameters }
Throughout the network, all the convolution layers use the $3{\times}3$ kernel with the LeakyReLU activation~\cite{xu2015empirical} without bias and have the same number of output channels as the input. For example, if an input of the convolution layer is of size 12 $\times$ Height $\times$ Weight, the output size will be the same and the number of learnable parameters on this layer is $3 \times 3 \times 12 \times 12=1296$ parameters, following the structure of standard convolutional layers~\cite{simonyan2015very}. The module \textit{Group PAConv} consists of  12 independent sub-modules where each has 144 learnable parameters from 4 convolutions whose input and output channels are 2 (i.e., $3 \times 3 \times 2 \times 2=144$ parameters). Therefore, each stage contains $14688$ learnable parameters, except for the last one which has  $9072$ parameters.
Table~\ref{tab:architecture} summarizes the network operations together with the corresponding number of learnable parameters, when considering $K=4$ stages.   
\begin{table}[ht]
\ra{1.00}
\caption{Summary of the architecture with the output shape and the number of learnable parameters. Stage 2 and 3 have the same structure as Stage 1. Conv denotes the convolution layer, and H and W represent the height and width of an array, respectively.}
\centering \scriptsize
\begin{tabular}{@{}lll l r @{}}
\toprule
Stage           & Block              & Layers                 & Output size & Parameters \\ \midrule
1                  & Feature extract. & (3 Conv layers)     & H$\times$W$\times$12       & 3,888             \\
                   & Squeeze            & PAConv (4 Conv layers) & H$\times$W$\times$12       & 5,184             \\
                   &                           & Argmax                 & H$\times$W$\times$1        &                  \\
                   & Expansion         & Feature extract. (3 Conv) & H$\times$W$\times$12        & 3,888               \\
                   &                    & Group PAConv         & H$\times$W$\times$12       & 1,728             \\
                   &                    & Convex combination       & H$\times$W$\times$12       &                  \\  \hline
2                  &                    &                        & H$\times$W$\times$12       & 14,688            \\ \hline
3                  &                    &                        & H$\times$W$\times$12       & 14,688            \\  \hline
4 		& Feature extract. & (3 Conv layers) & H$\times$W$\times$12       & 3,888             \\
                   & Squeeze            & PAConv (4 Conv layers) & H$\times$W$\times$12       & 5,184             \\
                   &                    & Argmax                 & H$\times$W$\times$1      &                 \\
\hline
\hline
 & Total &  &  &   53,136 \\
\bottomrule
\end{tabular}
\label{tab:architecture}
\end{table}

\subsubsection{Property of the network }
Interestingly, the final depth value of the proposed network is bounded pixelwise by the initial estimates of the multiscale depths.  To state formally, consider  the values of the multiscale depths $d_n^{(\ell)}, \forall \ell$ and the squeezed depth $x_n$ at the first stage. Since the squeeze block chooses an element among $L$ elements of $d_n^{(\ell)}$ with $\ell \in \{1,\cdots,L\}$, it holds that
\begin{equation} \label{eq:min_l}
\min \{ d_n^{(1)}, \cdots, d_n^{(\ell)} \} \leq x_n \leq \max \{ d_n^{(1)}, \cdots, d_n^{(\ell)} \}.
\end{equation}
Meanwhile, the expanded depths denoted by $\overline d^{(\ell)}_n$ are a convex combination of $d_n^{(\ell)}$ and $x_n$ with normalized weights in Eq.~\eqref{eq:overline d}, so we have
%
%
\begin{equation} \label{eq:min d_n^l}
 \min \{ d_n^{(\ell)}, x_n \} \leq \overline d^{(\ell)}_n \leq \max \{ d_n^{(\ell)}, x_n \}.
\end{equation}
Combining~\eqref{eq:min_l} and \eqref{eq:min d_n^l}, the expanded depths are bounded pixelwise by the initial multiscale depths
\begin{equation} \label{eq:min_l d_n^l}
\min \{ d_n^{(1)}, \cdots, d_n^{(\ell)} \} \leq \overline d^{(\ell)}_n \leq \max \{ d_n^{(1)}, \cdots, d_n^{(\ell)} \}.
\end{equation}

Since each stage has the same structure, this relation holds for the next stages and the final squeezed depth value has the same bound as in~\eqref{eq:min_l}. This property has pros and cons. We can predict the behaviour of the network, so that it will not produce some extreme depth values. On the other hand, the proposed network requires the range of initial multiscale depths to cover the underlying true depth for each pixel.

\subsection{Loss}

Motivated by the Laplace prior in~\eqref{eq:xndn}, we define the training loss for depths as the $\ell_1$-norm distance between the predicted depth and ground-truth depth. We additionally impose a constraint that the intermediate squeezed depths should be similar to the ground-truth depth $\boldsymbol {x}^*$ during training. The motivation for this constraint is twofold. It can prevent the neural network from losing key information in the initial stages and it can help avoid the vanishing gradient problem, by providing more paths in computational graphs for backpropagation. With the additional constraint, the training loss function $\mathcal L$ is defined as
\begin{equation} 
\mathcal L(\theta) = \sum_{k=1}^K \| \boldsymbol x^k(\theta) - \boldsymbol {x}^* \|_1,
\end{equation}
where $\theta$ denotes the neural network parameters, $\boldsymbol x^k$ represents the intermediate squeezed depth in the $k$ stage and $K$ is the total number of stages.

\subsection{Estimation of initial multiscale depths} \label{sec:estimation}

As a reminder, the input of the proposed network is a tuple of multiscale depths, rather than the large volume histogram data. From the histogram data, we aim at extracting initial multiscale depths without losing important information, while providing several depth values to cover the true one. For this goal, we consider several 3D low-pass filtered Lidar histograms as summarized in Table~\ref{tab:procedure}. We first apply the cross correlation to the original Lidar data with the system IRF. To this cross correlated data, we apply the 3D convolution with the uniform filters with the sizes of $7{\times}7{\times}7$ and $13{\times}13{\times}13$, generating two additional histograms. Each of the three histograms is then spatially downsampled with 4 different kernel sizes. This results in 12 filtered histograms in total, where we locate the main peak's position in each filtered histogram and for each pixel, to obtain initial multiscale depths. Note that we can exploit the separability of uniform filters for efficient computation. Note also that the actual IRF could be used and that we do not impose any constraints on the IRF shape.
It is worth mentioning that previous deep learning models~\cite{lindell2018singlephoton,peng2020photonefficient} do not account for a known system IRF in their architectures, but might learn it implicitly during training.
\begin{table}[ht]  \centering \scriptsize
\caption{Procedure to estimate initial multiscale depths.}
\begin{tabular}{cccccc}  \toprule
\multirow{2}{*}{1. Filter Lidar data}  & 2. Downsample & 3. Estimate \\
& with the filter size & initial depths \\
\midrule
\multirow{4}{*}{\begin{tabular}{l} $\boldsymbol Y \ast g$ (cross correlation) \\ $\boldsymbol Y$: Lidar data \\  $g$: System IRF \phantom{ filter aa} \end{tabular} } 
 	& $1{\times}1$  &  $\boldsymbol d^{\mathrm{ML}(1)}$ \\
 	& $3{\times}3$  &  $\boldsymbol d^{\mathrm{ML}(2)}$ \\
 	& $7{\times}7$ &   $\boldsymbol d^{\mathrm{ML}(3)}$  \\
 	& $13{\times}13$  & $\boldsymbol d^{\mathrm{ML}(4)}$  \\ \midrule
 	
\multirow{4}{*}{\begin{tabular}{l} $\boldsymbol Y \ast g \ast \omega_1$\\$\omega_1$: Uniform filter kernel\\ \phantom{w11}(size: $7{\times}7{\times}7$)\end{tabular} } 
  & $1{\times}1$ & $\boldsymbol d^{\mathrm{ML}(5)}$    \\
 	& $3{\times}3$ & $\boldsymbol d^{\mathrm{ML}(6)}$  \\
 	& $7{\times}7$ & $\boldsymbol d^{\mathrm{ML}(7)}$ \\
 	& $13{\times}13$ & $\boldsymbol d^{\mathrm{ML}(8)}$ \\ \midrule
 
\multirow{4}{*}{\begin{tabular}{l} $\boldsymbol Y \ast g \ast \omega_2$\\$\omega_2$: Uniform filter kernel\\\phantom{w11}(size: $13{\times}13{\times}13$)\end{tabular} } 
   & $1{\times}1$ & $\boldsymbol d^{\mathrm{ML}(9)}$  \\
 	& $3{\times}3$  & $\boldsymbol d^{\mathrm{ML}(10)}$ \\
 	& $7{\times}7$  &   $\boldsymbol d^{\mathrm{ML}(11)}$ \\
 	& $13{\times}13$ & $\boldsymbol d^{\mathrm{ML}(12)}$  \\
\bottomrule
\end{tabular}
\label{tab:procedure}
\end{table}
\subsection{Training procedures} \label{sec:training}

To train the neural network, we generate synthetic data by simulating SPAD measurements with $T=1024$ time bins, using the Poisson observation model in \eqref{eq:snt}. We choose   9 scenes from the Middlebury stereo dataset~\cite{hirschmuller2007evaluation} (with image sizes $555{\times}650$) and   21 scenes from the Sintel stereo dataset~\cite{butler2012naturalistic} (with   image sizes $436{\times}1024$) for the training dataset and 2 scenes from~\cite{butler2012naturalistic} for the validation set. To make our network robust to different noise levels, we consider different scenarios based on the average number of  Photons-Per-Pixel (PPP); and the average Signal-to-Background Ratio (SBR),  defined as
$$
\text{PPP}=\frac{1}{N}\sum_{n=1}^N \left(r_n + b_{n} T\right), \,\, \text{SBR}=\frac{\sum_{n=1}^N r_n} { \sum_{n=1}^N  b_{n} T }.
$$
We consider 4 cases: (PPP=1, SBR=1), (PPP=1, SBR=64), (PPP=64, SBR=1), (PPP=64, SBR=64). To save the GPU memory during training, we extract patches of size $256{\times}256$ with stride $48$, rather than processing the original images. We implement our model in PyTorch and use ADAM~\cite{kingma2015adam} as an optimizer with the default hyperparameter ($\beta_1$=$0.9$, $\beta_2$=$0.999$) and the batchsize 16. We train the model for 200 epochs with the initial learning rate 0.0001 which is reduced by half at epoch 100. The training was performed on a Linux server with a NVIDIA RTX 3090 GPU, which takes about 9 hours.

\section{Experimental results} \label{sec:experimental}

\begin{figure}[t]
\includegraphics[width=250pt]{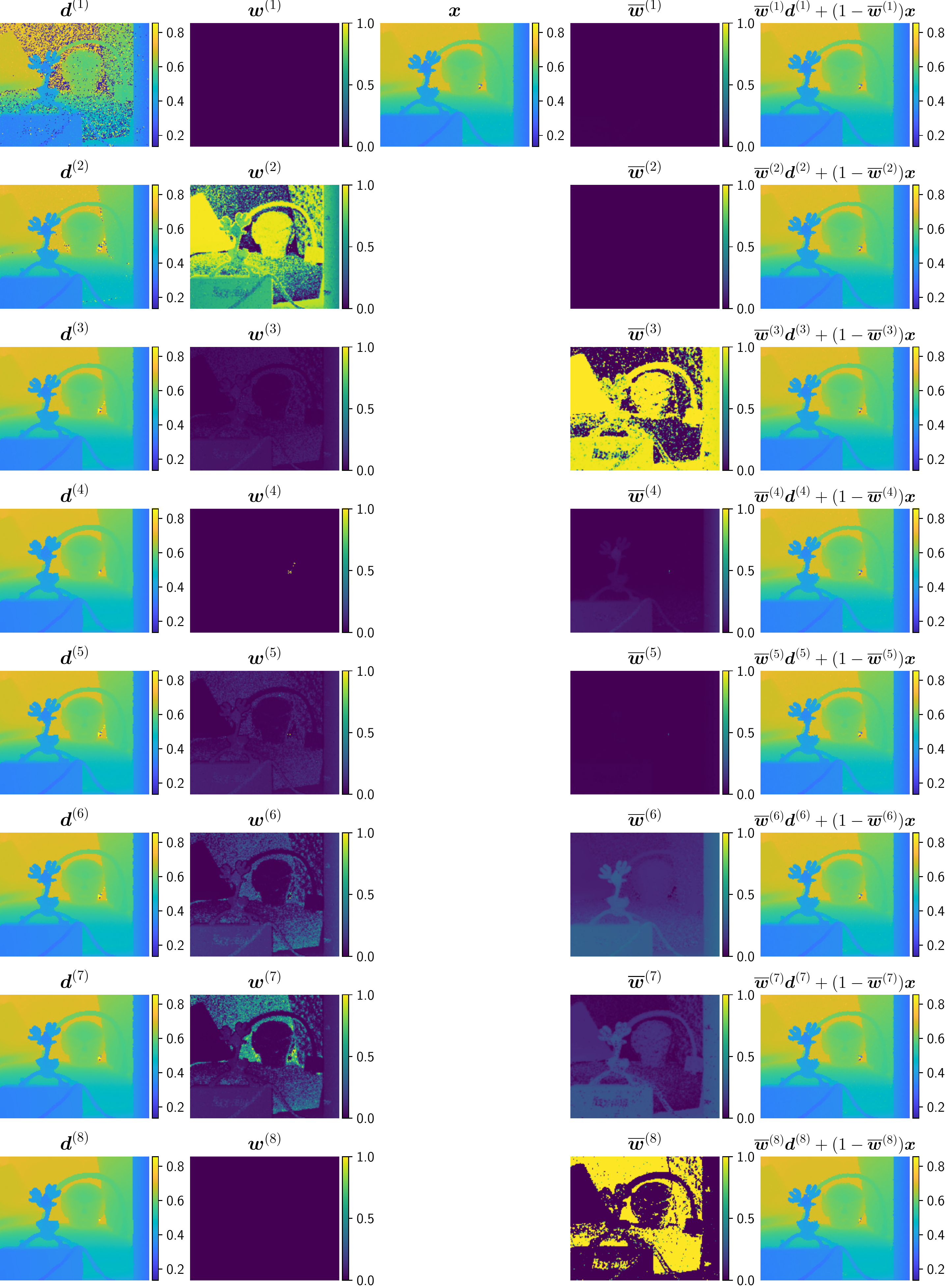}
\caption{Visualization of the internal outputs in the first stage. For the purpose of presentation, only eight scales ($\ell=1,\cdots,8$) are visualized. [1st column] shows the initial multiscale depths $(\boldsymbol d)$ and [2nd column] their corresponding attention weights $(\boldsymbol w)$ in the squeeze block. [3rd column] shows the squeezed depth $(\boldsymbol x)$, [4th column] the attention weights in the expansion block and [5th column] the expanded multiscale depths computed by Eq.~\eqref{eq:overline d}.}
\label{fig:visualization}
\end{figure}

\begin{figure}[t]
\centering \includegraphics[width=250pt]{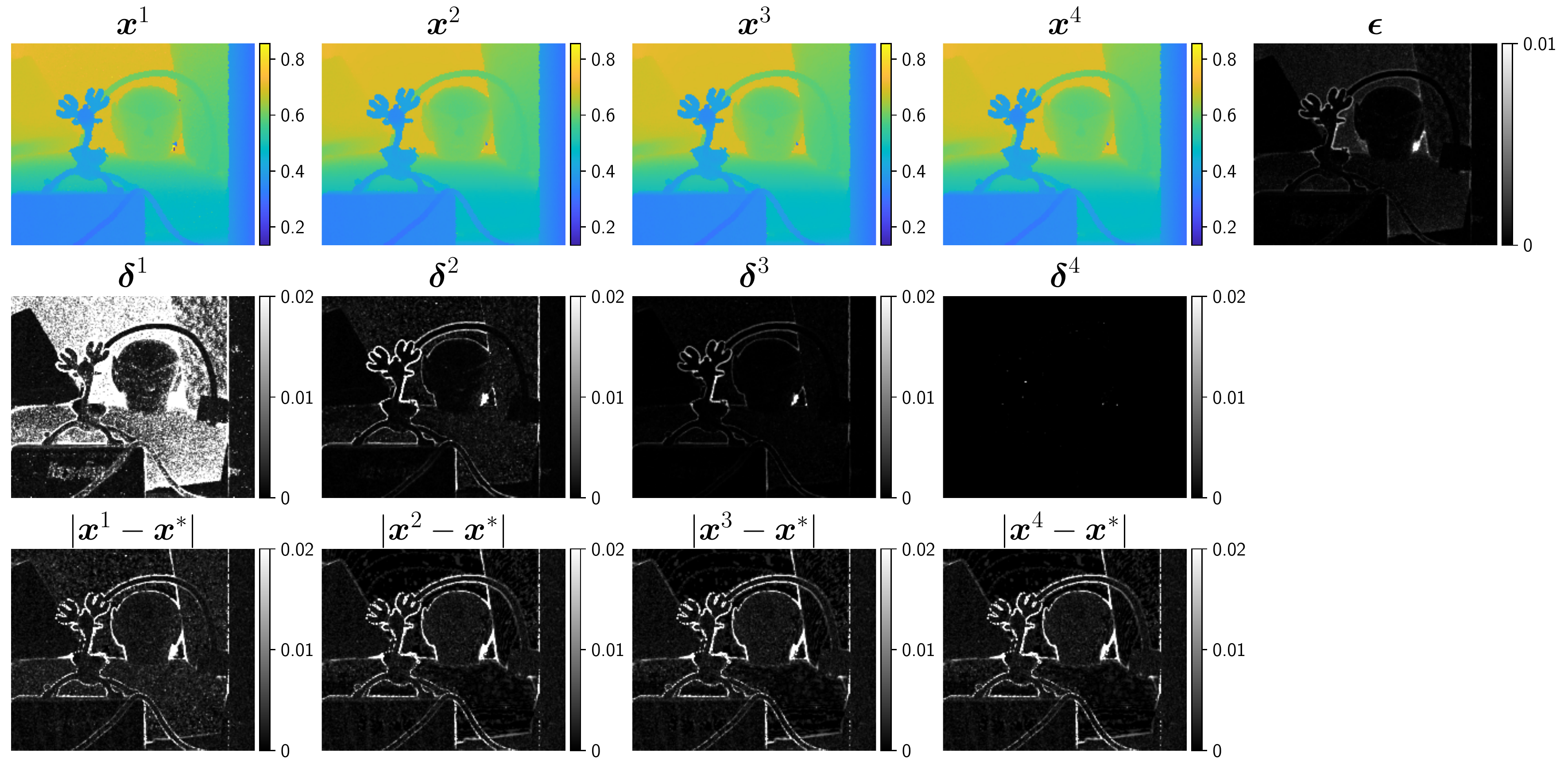}
\caption{Intermediate squeezed depths along the four stages (top), the difference between the multiscale depths and the corresponding squeezed depth in each stage (middle) and the errors between the squeezed depths and the ground-truth (bottom). The last column shows our estimated uncertainty map.}
\label{fig:visualization2}
\end{figure}

In this section, we perform the experiments to analyze our model and show the relative advantages over other reconstruction methods on synthetic datasets as well as real datasets.

\subsection{Analysis of the network}

\myhead{Test dataset.} For the test data, we simulated Lidar data with    $T=1024$ time bins, from two scenes of \textit{Art} ($555{\times}695$) and \textit{Reindeer} ($555{\times}671$) in the Middlebury stereo dataset~\cite{hirschmuller2007evaluation} which did not belong to our training sets. The reference depth   and   reflectivity maps of these two scenes are visualized in the first column of Figs.~\ref{fig:depth1} and~\ref{fig:depth2}, respectively. In particular, the Reindeer scene contains extremely low-photon regions which will challenge the reconstruction algorithms. Note that these test data are larger than the data reported in previous deep learning works~\cite{lindell2018singlephoton,peng2020photonefficient}.
    \def\scene{Art}
    \def\fh{39pt}
    \def\fw{49pt}
    \def\seqa{16.0_4.0}
    \def\seqb{4.0_1.0}
    \def\seqc{1.0_0.25}
    \def\pppa{16} \def\sbra{4}
    \def\pppb{4} \def\sbrb{1}
    \def\pppc{1} \def\sbrc{0.25}
    \begin{figure*}[!ht]
    \centering
\begin{tabular}{c@{ }c@{ }c@{ }c@{ }c@{ }c@{ }c@{ }c@{ }c@{ }c@{ }c@{ }c}
& \rotatebox[origin=l]{90}{\small\parbox{1.3cm}{\scriptsize$\,$PPP = \pppa\\SBR = \sbra}}&
\includegraphics[totalheight=\fh]{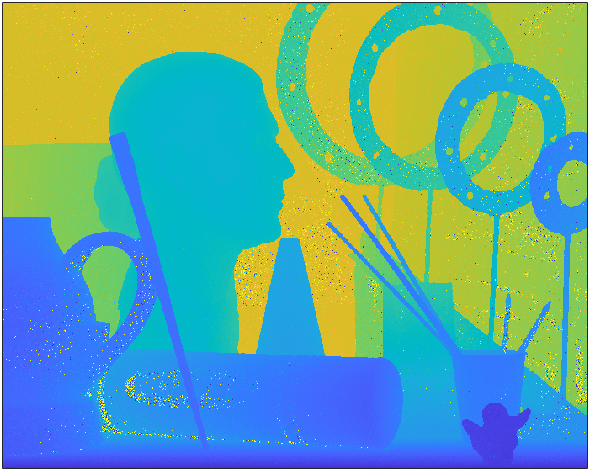} &
\includegraphics[totalheight=\fh]{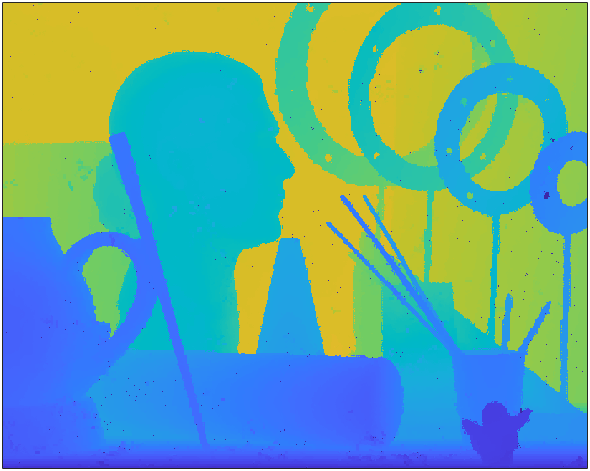} & 
\includegraphics[totalheight=\fh]{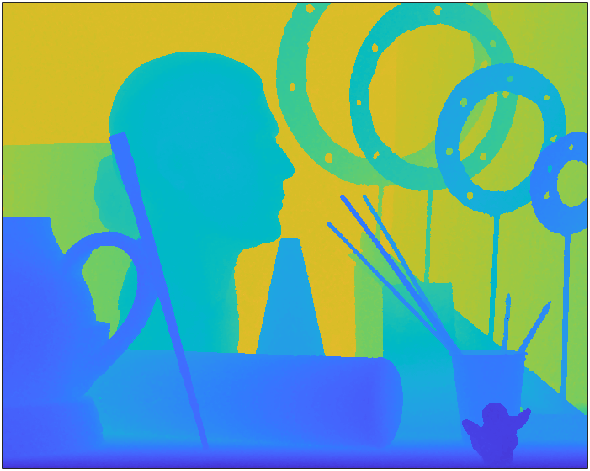} & 
\includegraphics[totalheight=\fh]{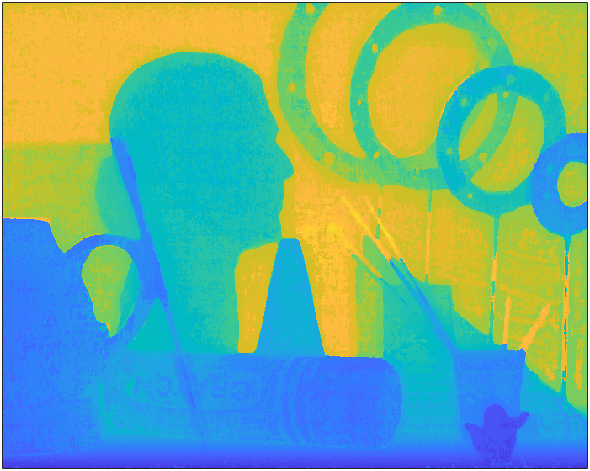} & 
\includegraphics[totalheight=\fh]{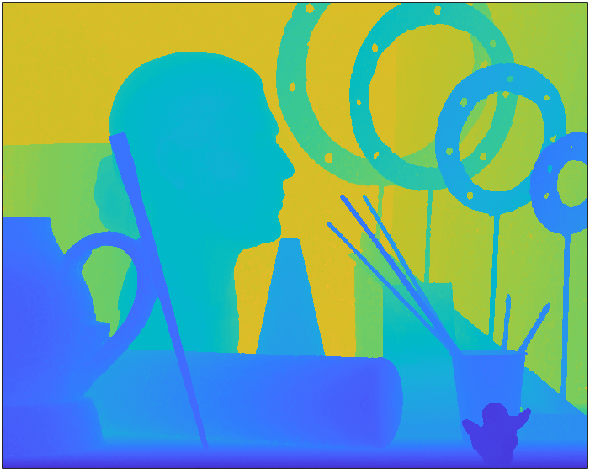}&
\includegraphics[totalheight=\fh]{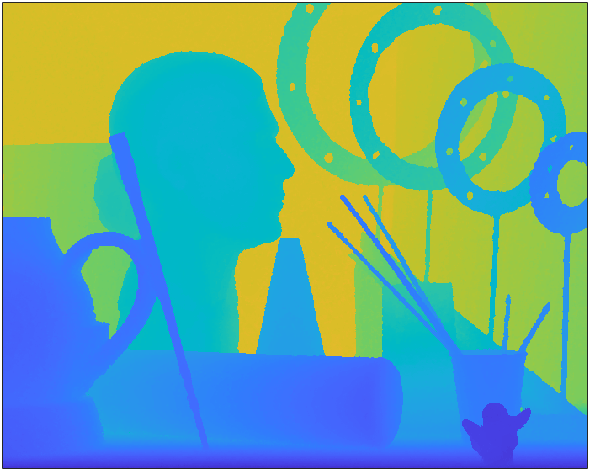} & \includegraphics[totalheight=\fh]{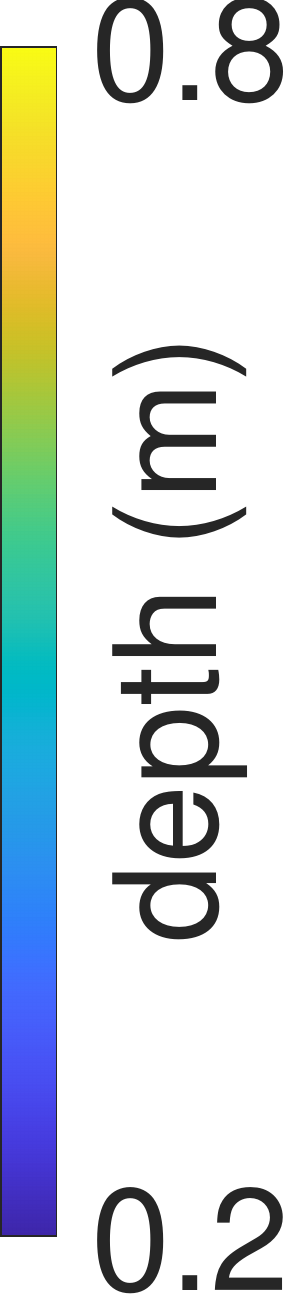} & \includegraphics[totalheight=\fh]{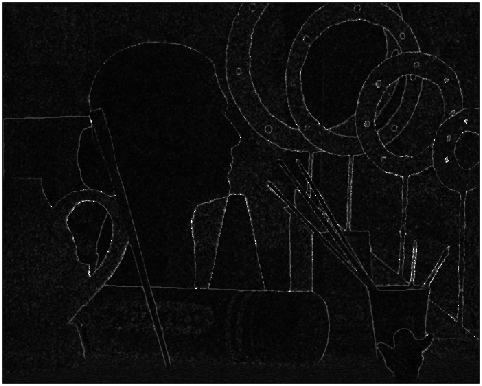}  &\includegraphics[totalheight=\fh]{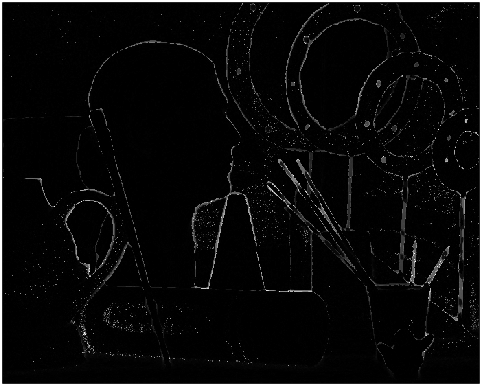}  &\includegraphics[totalheight=\fh]{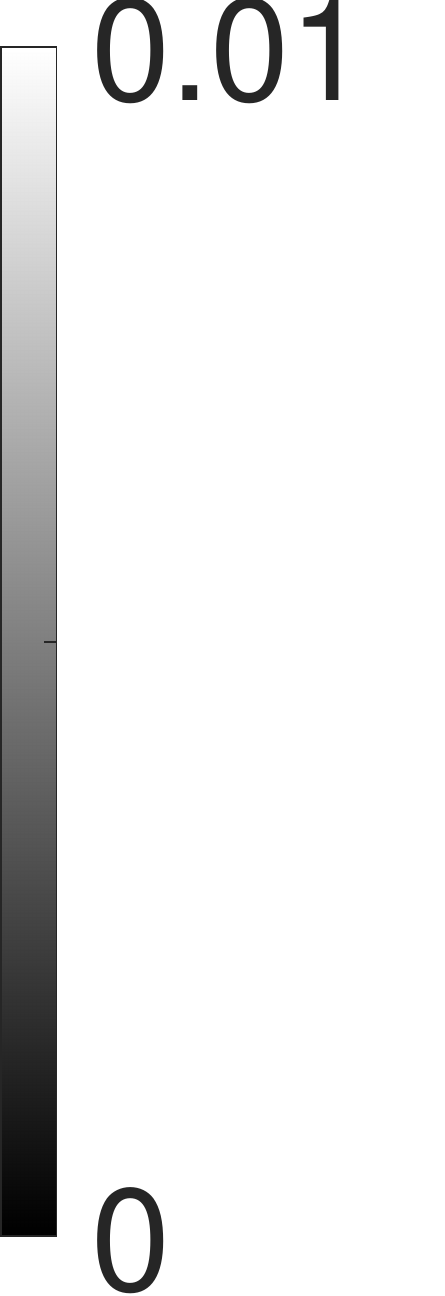}  \\

\includegraphics[totalheight=\fh]{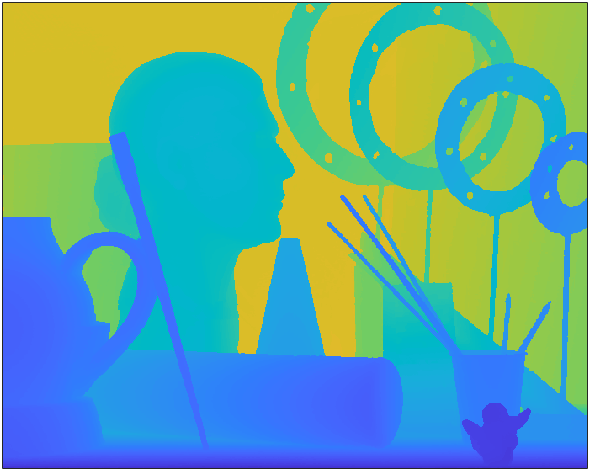}  &
\rotatebox[origin=l]{90}{\small\parbox{1.3cm}{\scriptsize$\,$PPP = \pppb\\SBR = \sbrb}}&
\includegraphics[totalheight=\fh]{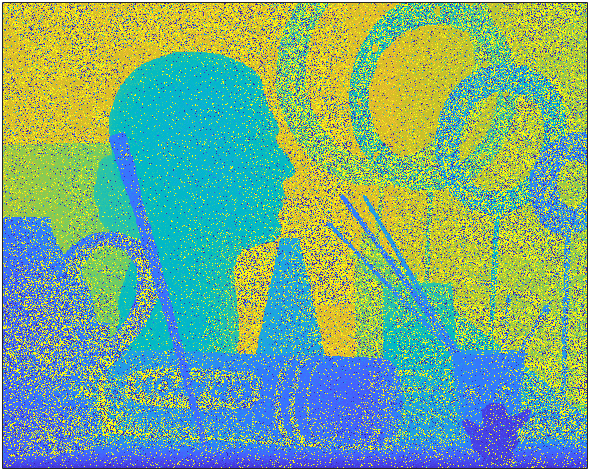} &
\includegraphics[totalheight=\fh]{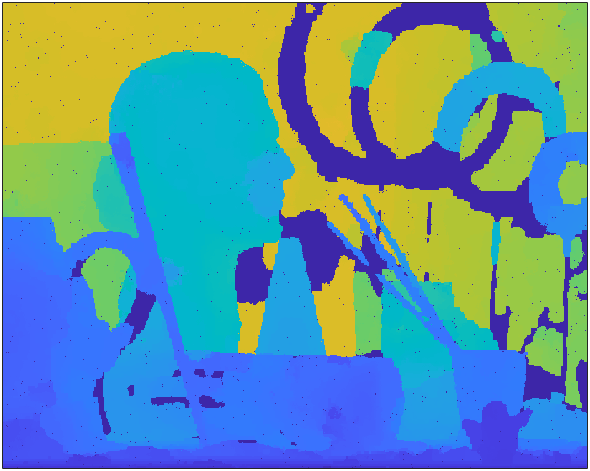} & 
\includegraphics[totalheight=\fh]{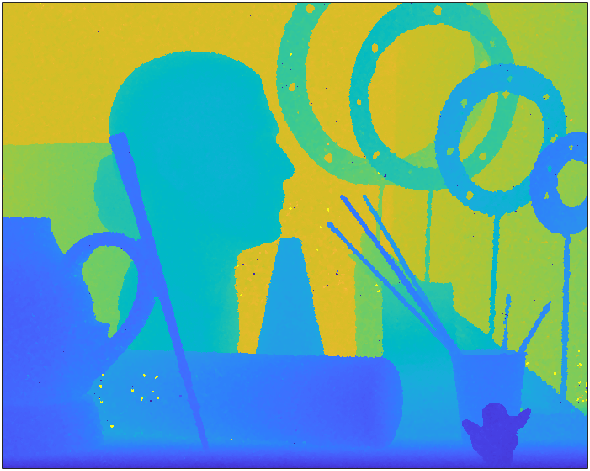} & 
\includegraphics[totalheight=\fh]{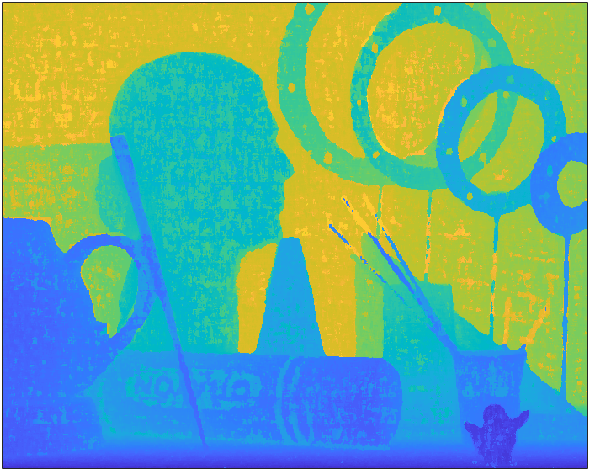} & \includegraphics[totalheight=\fh]{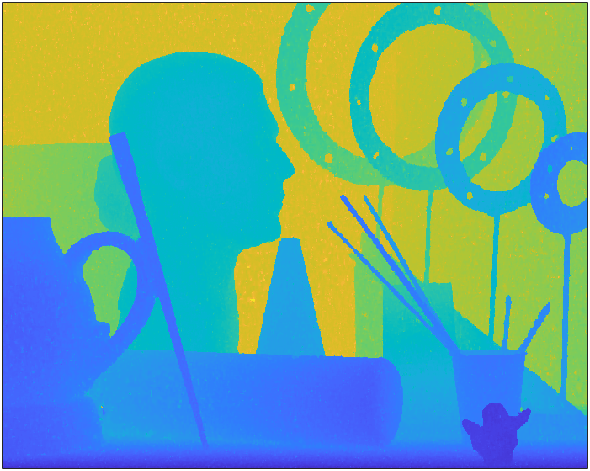} &
\includegraphics[totalheight=\fh]{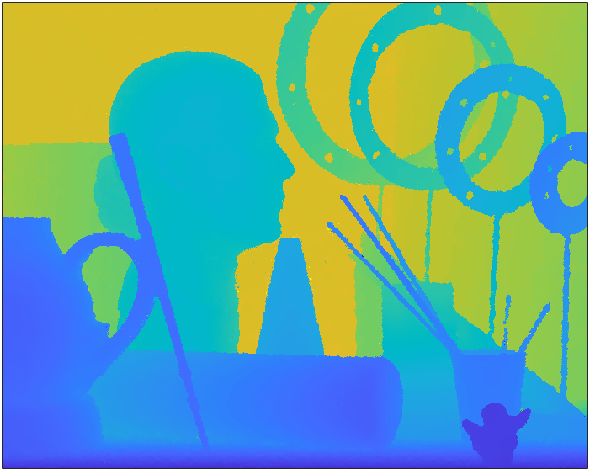}  &\includegraphics[totalheight=\fh]{exp11/colorbar} & \includegraphics[totalheight=\fh]{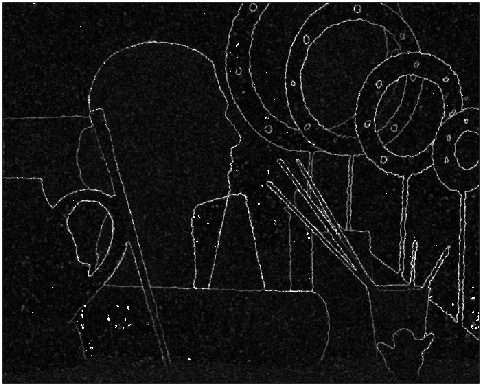}   & \includegraphics[totalheight=\fh]{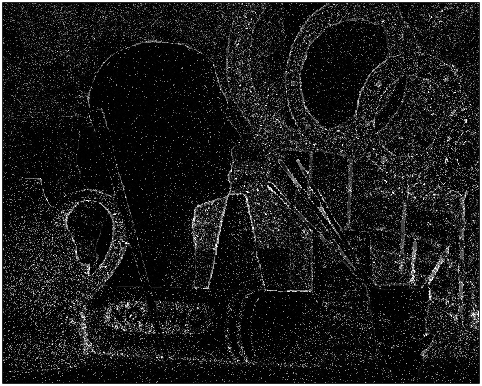} & \includegraphics[totalheight=\fh]{exp11/colorbar_uncertainty}  \\ 

\includegraphics[width=\fw,totalheight=\fh]{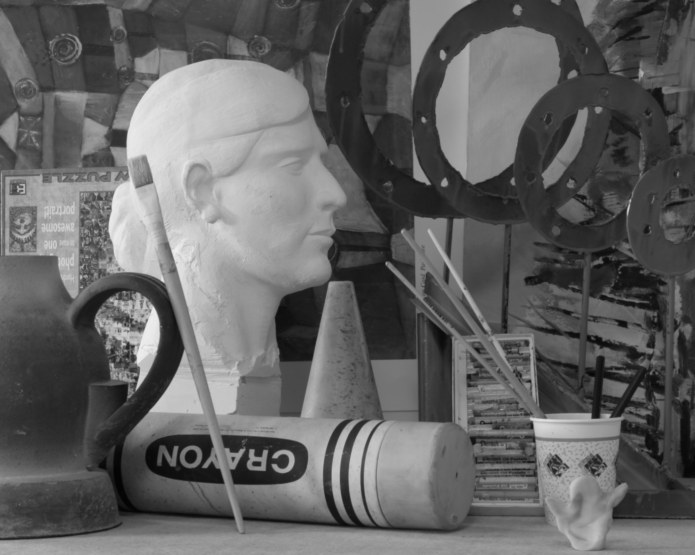}  &
 \rotatebox[origin=l]{90}{\small\parbox{1.3cm}{\scriptsize$\,$PPP = \pppc\\SBR = \sbrc}}&
\includegraphics[totalheight=\fh]{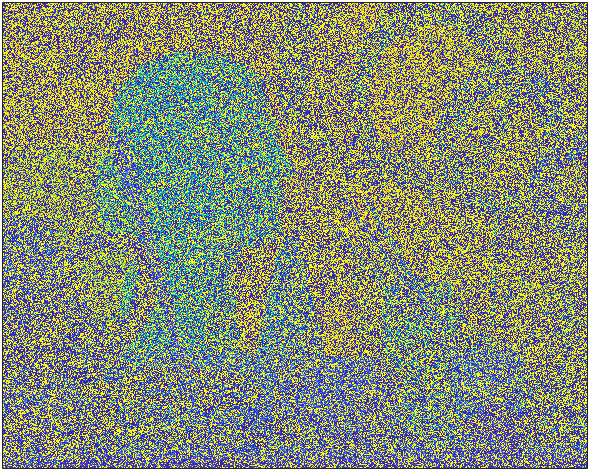} &
\includegraphics[totalheight=\fh]{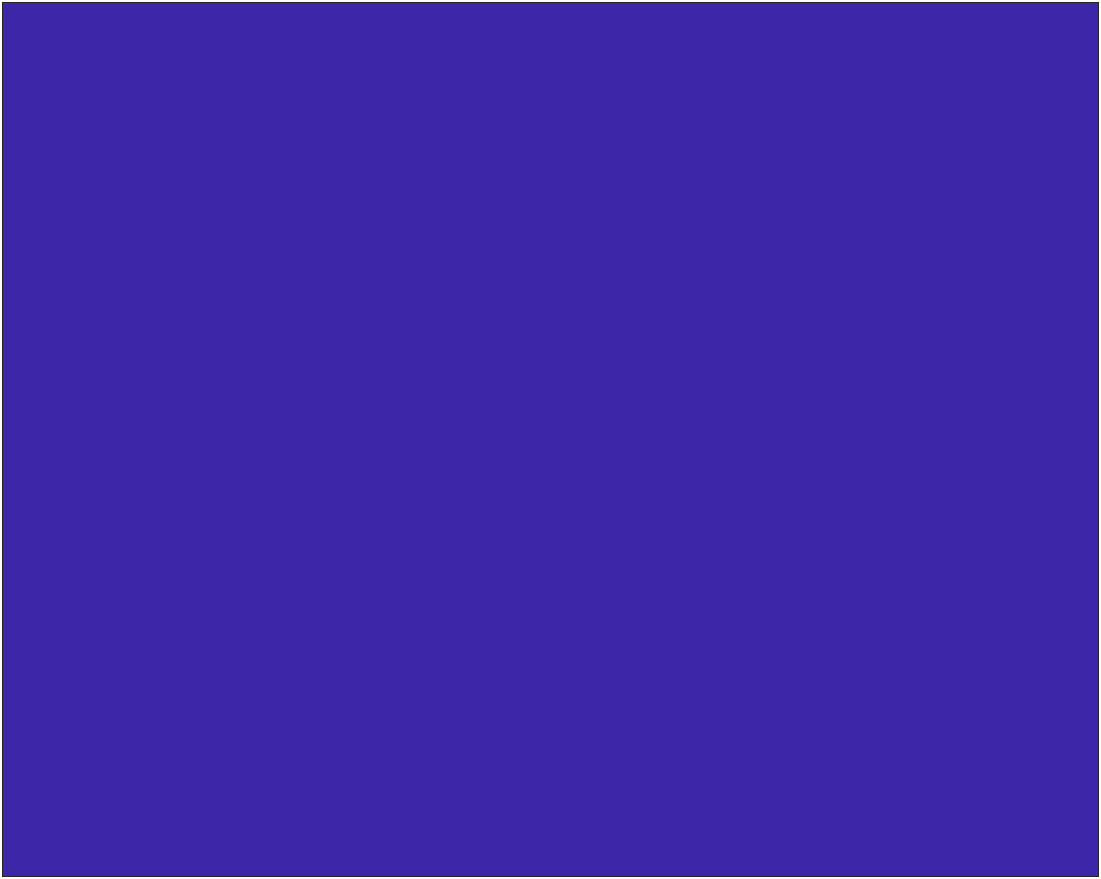} & 
\includegraphics[totalheight=\fh]{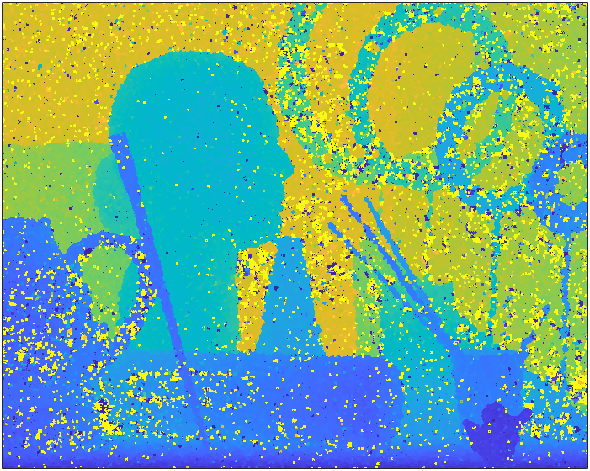} & 
\includegraphics[totalheight=\fh]{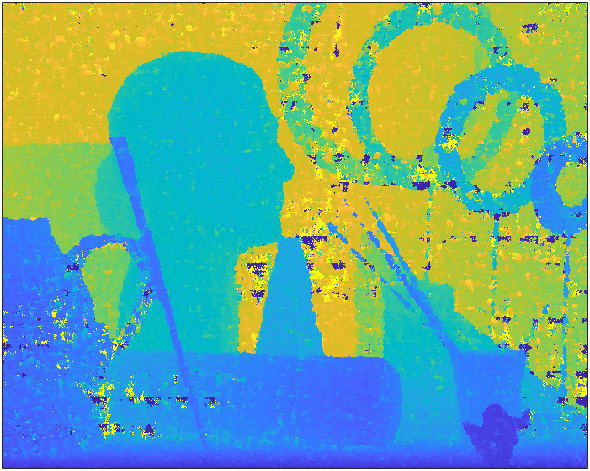} & \includegraphics[totalheight=\fh]{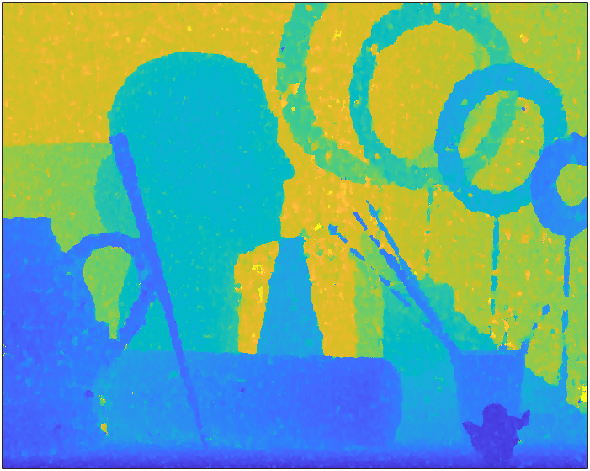} &
\includegraphics[totalheight=\fh]{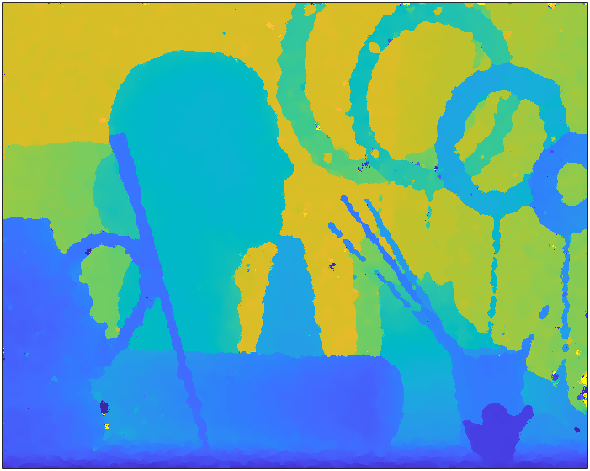}  &\includegraphics[totalheight=\fh]{exp11/colorbar} &\includegraphics[totalheight=\fh]{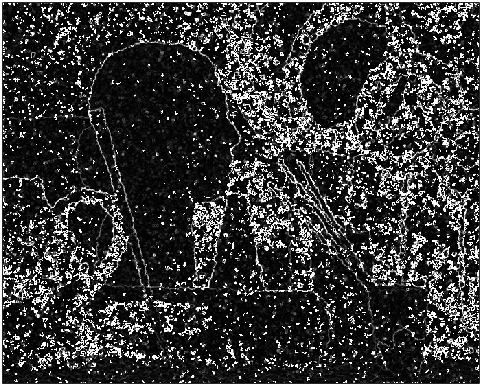}   & \includegraphics[totalheight=\fh]{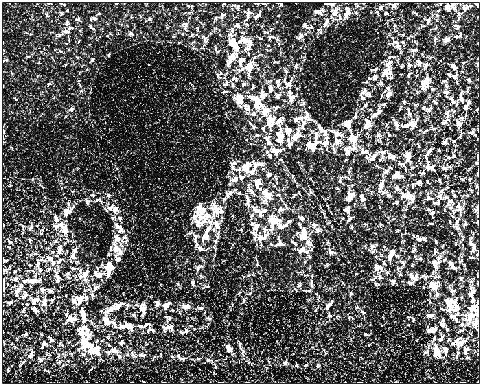}  &\includegraphics[totalheight=\fh]{exp11/colorbar_uncertainty} \\
\multirow{2}{*}{Reference} & & \multirow{2}{*}{Classic} & \multirow{2}{*}{Manipop} & \multirow{2}{*}{Halimi} & \multirow{2}{*}{Lindell} & \multirow{2}{*}{Peng} & \multirow{2}{*}{Proposed} && Halimi &Proposed& \\
&&&&&&&&& \multicolumn{2}{c}{Uncertainty}
\end{tabular}
\caption{Reconstructed depth maps with different PPP and SBR levels on the \scene{} scene. The first column shows the reference depth map (middle) and the reflectivity image (bottom). The last two columns show the estimated uncertainty maps by Halimi and the proposed method.}
    \label{fig:depth1}
    \end{figure*}
    \def\fw{60pt}
    \def\fhpc{50pt}
    \def\ftype{pc_wo_axis}
    \begin{figure*}[!ht]
    \centering
    \begin{tabular}{c@{ }c@{ }c@{ }c@{ }c@{ }c@{ }c@{ }c}
     & \rotatebox[origin=l]{90}{\small\parbox{1.5cm}{\scriptsize$\,$PPP $=\pppa$\\SBR $=\sbra$}}&
    \includegraphics[width=\fw,totalheight=\fhpc]{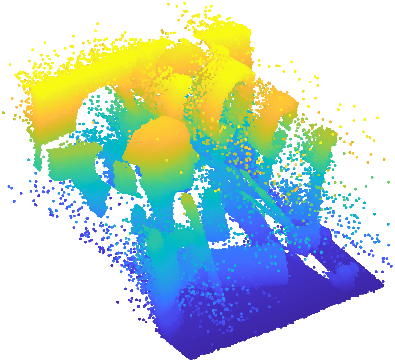} &
    \includegraphics[width=\fw,totalheight=\fhpc]{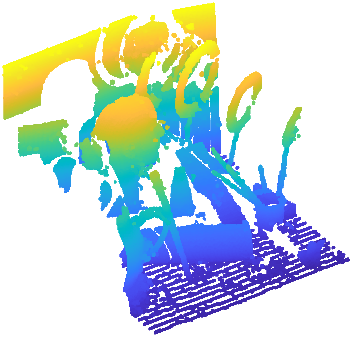} &
    \includegraphics[width=\fw,totalheight=\fhpc]{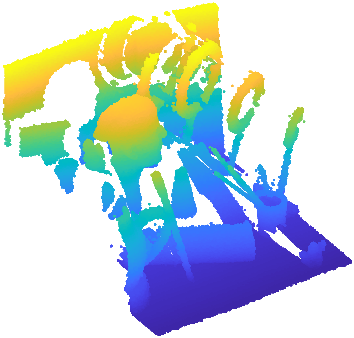} &
    \includegraphics[width=\fw,totalheight=\fhpc]{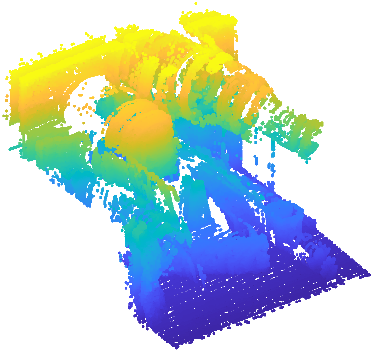} & \includegraphics[width=\fw,totalheight=\fhpc]{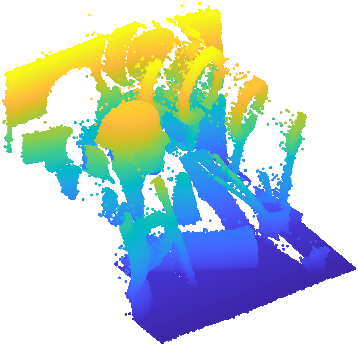} & \includegraphics[width=\fw,totalheight=\fhpc]{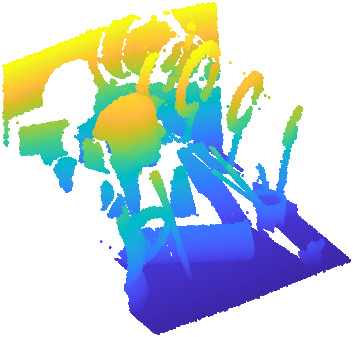} \\

    \includegraphics[width=\fw,totalheight=\fhpc]{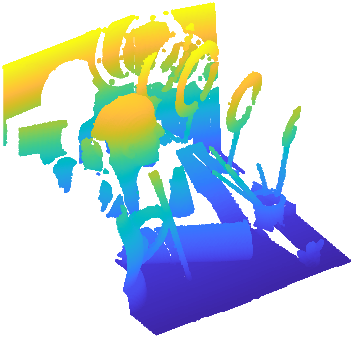}  &
    \rotatebox[origin=l]{90}{\small\parbox{1.5cm}{\scriptsize$\,$PPP $=\pppb$\\SBR $=\sbrb$}} &
    \includegraphics[width=\fw,totalheight=\fhpc]{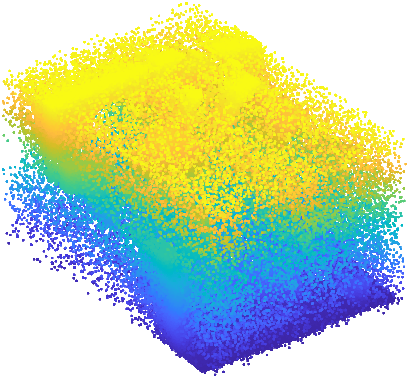} &
    \includegraphics[width=\fw,totalheight=\fhpc]{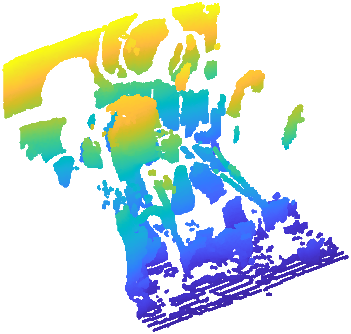} &
    \includegraphics[width=\fw,totalheight=\fhpc]{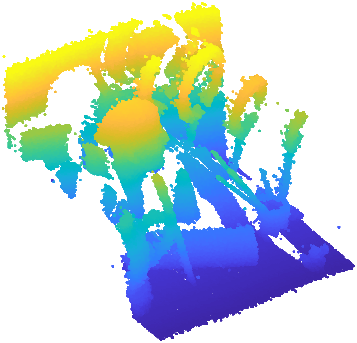} &
    \includegraphics[width=\fw,totalheight=\fhpc]{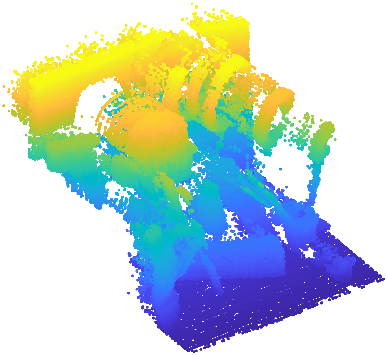} & \includegraphics[width=\fw,totalheight=\fhpc]{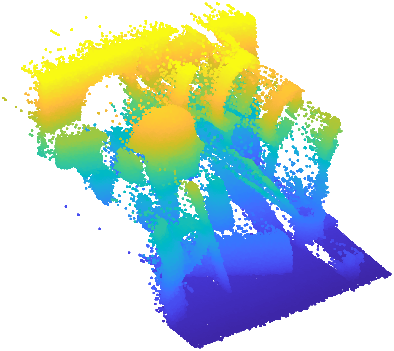} & \includegraphics[width=\fw,totalheight=\fhpc]{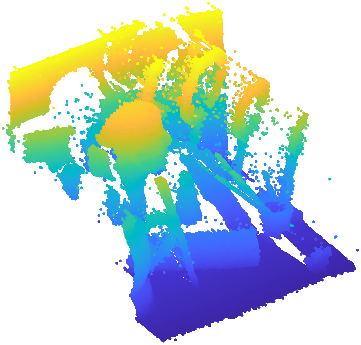} \\

 &
   \rotatebox[origin=l]{90}{\small\parbox{1.5cm}{\scriptsize$\,$PPP $=\pppc$\\SBR $=\sbrc$}}&
   \includegraphics[width=\fw,totalheight=\fhpc]{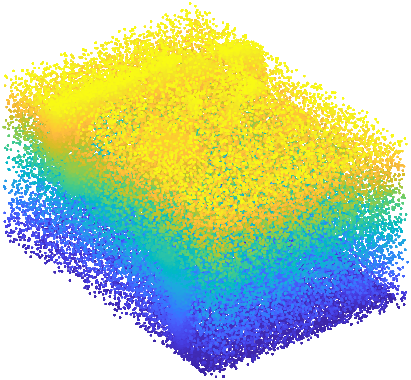} &
   & 
    \includegraphics[width=\fw,totalheight=\fhpc]{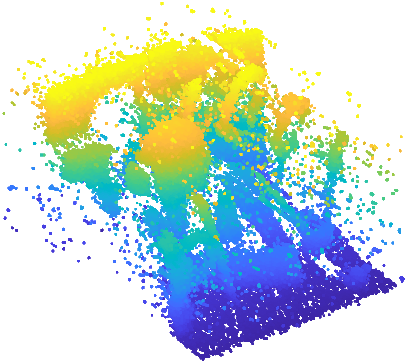} &
    \includegraphics[width=\fw,totalheight=\fhpc]{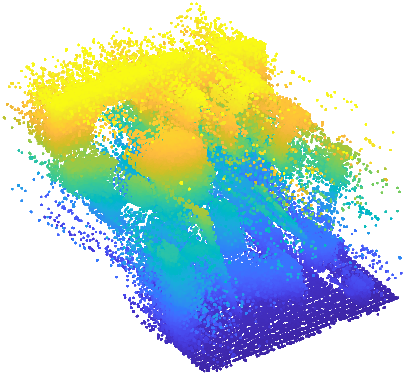} & \includegraphics[width=\fw,totalheight=\fhpc]{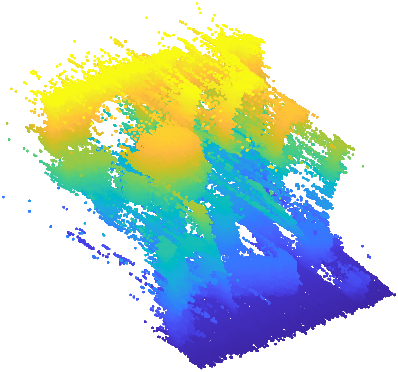} & \includegraphics[width=\fw,totalheight=\fhpc]{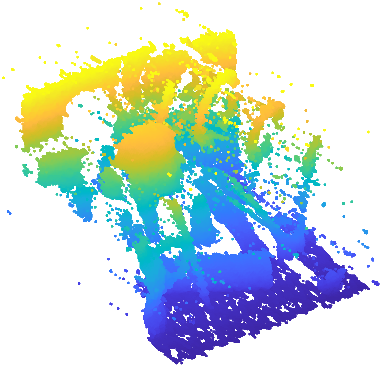} \\
    Reference & & Classic & Manipop & Halimi & Lindell & Peng & Proposed 
    \end{tabular}
    \caption{Point cloud representation of reconstruction results on the \scene{} scene. The first column shows the reference point cloud.}    
    \label{fig:pc1}
    \end{figure*}

\myhead{Interpretability.} Thanks to our unrolling strategy, we can interpret our neural network via the connection to the underlying Bayesian method. We first inspect whether the first stage  can successfully discard outliers. Fig.~\ref{fig:visualization} visualizes the outputs of the internal blocks in the first stage. Each pixel of the squeezed depth $\boldsymbol x$ (3rd column) is obtained using \eqref{eq:argmax}, where the multiscale depths and weights are represented in the 1st and 2nd  columns, respectively.  As shown in the first row, the first scale depth $\boldsymbol d^{(1)}$ shows many outliers, which leads to zero values in $\boldsymbol w^{(1)}$. On the other hand, the second scale depth $\boldsymbol d^{(2)}$ contains important features with less noise and its attention weight $\boldsymbol w^{(2)}$ contains many high values. We still observe noise in $\boldsymbol d^{(2)}$ especially around the low-photon regions. In such areas, $\boldsymbol d^{(4)}$ and $\boldsymbol d^{(7)}$ show more smoothed depth values and $\boldsymbol w^{(4)}$, $\boldsymbol w^{(7)}$ receive  higher attention weights. In this way, the proposed network can successfully remove noise and discard many outliers in the first stage. The 4th column shows the attention weights for the expansion block where $\overline{\boldsymbol w}^{(1)}$ has only zero values, which indicates that the first scale depth $\boldsymbol d^{(1)}$ will be discarded in the next stage.

We now investigate how the squeezed depths improve along the stages. To quantify the change within each stage, we define the difference $\boldsymbol \delta^k$ in the $k$th stage between the multiscale depths and the corresponding squeezed depth, for each pixel $n$, as follows:
\begin{equation} \label{eq:delta}
\delta^k_n = \frac{1}{L+2} \sum_{\ell=1}^L | x^k_n - d_n^{k,(\ell)}|,
\end{equation}
where $d_n^{k,(\ell)}$ is the multiscale depth value in the $k$th        stage. A small value of $\boldsymbol \delta$ would indicate a small improvement so that we may not need a further stage. We also define the uncertainty $\boldsymbol \epsilon$ of our final depth map $\boldsymbol x^K$ via the connection to the mode of the depth variance~\eqref{eq:epsilon_n} in the underlying method with $\bar N=1$, as follows:
\begin{equation} \label{eq:epsilon_ours} \small
\epsilon_n = \frac{1}{K-1} \sum_{k=1}^{K-1} \frac{\mathcal C^k_n + \beta_d }{L + 2 + \alpha_d}, \,\, \mathcal C^k_n = \sum_{\ell=1}^L \overline{\overline{w}}_n^{k,(\ell)} |d^{k,(\ell)}_n - x^K_n |,
\end{equation}
where $\overline{\overline{\boldsymbol w}}^k$ is the softmax-normalized version of $1 -  \overline{\boldsymbol w}^k$, ensuring $\sum_{\ell=1}^L \overline{\overline{w}}_n^{k,(\ell)}=1$. The weights $\overline{\overline{\boldsymbol w}}$   play a similar role to the guidance weights in~\eqref{eq:dnargmin}, and we set the hyperparameters $\alpha_d,\beta_d$ 
to small values to obtain a non-informative prior.
Fig.~\ref{fig:visualization2} shows the intermediate squeezed depths $\boldsymbol x^k$ for each stage $k$ and $\boldsymbol \delta^k$ which decreases along the stages. The last row shows the errors between the squeezed depths and the ground-truth depth map $\boldsymbol x^*$. In the first stage, the errors appear on the background due to outliers, but such errors decrease along the stages. The last column shows the estimated uncertainty map $\boldsymbol \epsilon$. It indicates high uncertainty around object edges and areas with low reflectivity.

%
    \def\scene{Reindeer}
    \def\fw{47pt}
    \begin{figure*}[!ht]
    
    \label{fig:depth2}
    \end{figure*}
    \def\fw{60pt}
    \begin{figure*}[!ht]
        
    \label{fig:pc2}
    \end{figure*}

\myhead{Evaluation metrics.} To analyze our model quantitatively, we use three evaluation metrics. We employ a standard metric, Depth Absolute Error (DAE) defined as $\operatorname{DAE}\,(\boldsymbol x, \boldsymbol {x^*}) = \frac{1}{N} \| \boldsymbol x - \boldsymbol {x}^* \|_1$, where $N$ is the number of pixels, which is useful for measuring the overall disparity quality. To better evaluate surface boundaries, we use an additional metric called Soft Edge Error (SEE)~\cite{chen2019oversmoothing}, which measures the local error only at the edges. Formally, it is defined as
$$\operatorname{SEE} \, ( \boldsymbol x, \boldsymbol {x}^* ) = \gamma \sum_{n \in Edge(\boldsymbol{x}^*)}   \min_{j \in \nu _n} | x_j - x_j^* |, $$
where $\nu_n$ is a $3\times 3$ local window around the $n$th pixel, $\gamma:=10 / |Edge(\boldsymbol {x}^*)|$ is a scale factor, and 
$Edge(\boldsymbol {x}^*)$ represents a set of edge locations in the ground-truth depth map $\boldsymbol{x}^*$ obtained using the Canny edge detector~\cite{canny1986computational}.
We also report the root mean square error: $\operatorname{RMSE}(\boldsymbol x, \boldsymbol {x}^*) = \sqrt{\| \boldsymbol x - \boldsymbol {x}^*\|_2^2 / N }$, as previously used in~\cite{lindell2018singlephoton,peng2020photonefficient}.

\myhead{Ablation study.} We study the effect of the number of stages $K$ and  scales $L$. As shown in Table~\ref{tab:effectKL}, we first fix $L=12$ and vary the number of stages $K$ from 2 to 5. We evaluate the performance on 98 different Lidar data with different levels of PPP and SBR both ranging from 0.25 to 1024. We note a decreasing error for increasing  number of stages,  but when $K=5$, the error is shown to increase possibly due to overfitting to our training data. For example, we observed that the case $K=5$ gives a worse performance than that of $K=4$ when SBR is less than 1 which did not belong to our tranining set. The number of stages affects the running time by a small margin, because most of the computational cost comes from generating the initial multiscale depths. Next, we test the effect of the total number of scales explained in Section~\ref{sec:estimation}. The number of scales $L$ affects the error, the number of parameters and the running time. To balance the trade-off between the performance and the network size, we choose $K=4$ and $L=12$ throughout the rest of the experiments.

\begin{table}[ht]  \centering
\caption{Effect of the number of stages $K$ and the number of scales $L$. The error DAE is presented with the mean values and the standard deviation, evaluated on 98 different Lidar data with different PPP and SBR on the Art and Reindeer scene.}
\begin{tabular}{c c crr}  \toprule
$K$ & $L$ & DAE & Parameters & Run time (sec) \\
\midrule
2 & \multirow[t]{4}{*}{12} 	& 0.0055 $\pm$ 0.0129  & 23,760 & \multirow{4}{*}{5.1 $\pm$ 0.1}  \\
3 & 	& 0.0046 $\pm$ 0.0119 & 38,448 &    \\
4 & 	& 0.0040 $\pm$ 0.0082  & 53,136 &   \\
5 & 	& 0.0043 $\pm$ 0.0112  & 67,824 &   \\
\midrule
\multirow[t]{3}{*}{4} & 4 & 0.0074 $\pm$ 0.0208  & 5,841 & 3.5  \\
 & 8	& 0.0053 $\pm$ 0.0151  & 24,768 & 4.0  \\
 & 12	& 0.0040 $\pm$ 0.0082  & 53,136 & 5.1  \\
\bottomrule
 \end{tabular}
\label{tab:effectKL}
\end{table}

\def\sbra{0.25}
\def\sbrb{4}
\begin{table*}[!ht]
\centering \ra{0.95}
\caption{Quantitative comparison on the Art and Reindeer scene with different levels of PPP and SBR.}
\begin{tabular}{@{}lccc c ccc c ccc c ccc @{}}\toprule
& \multicolumn{3}{c}{Art,\, SBR = \sbra} & & \multicolumn{3}{c}{Art,\, SBR = \sbrb} & &
 \multicolumn{3}{c}{Reindeer,\, SBR = \sbra} & & \multicolumn{3}{c}{Reindeer,\, SBR = \sbrb} \\ \cmidrule{2-4} \cmidrule{6-8} \cmidrule{10-12} \cmidrule{14-16}
& DAE & SEE & RMSE && DAE & SEE & RMSE & & DAE & SEE & RMSE && DAE & SEE & RMSE\\ \midrule
PPP = 1\\
Manipop& - & - & - &   & 0.1883 & 1.3787 & 0.3161 & & - & - & - &  & 0.2579 & 1.2064 & 0.3919 \\
Halimi & 0.2188 & 0.3269 & 0.5739 &  & 0.0206 & 0.0170 & 0.1422 &  & 0.1581 & 0.1550 & 0.4891 &  & 0.0110 & 0.0076 & 0.0874 \\
Lindell & 0.0489 & 0.0597 & 0.2040 &  & 0.0115 & 0.0172 & 0.0323 &  & 0.0916 & 0.0874 & 0.3112 &  & 0.0111 & 0.0308 & 0.0313 \\
Peng & 0.0170 & 0.0424 & \textbf{0.0722} &  & 0.0060 & 0.0130 & \textbf{0.0151} &  & 0.0132 & 0.0425 & \textbf{0.0371} &  & 0.0066 & 0.0115 & \textbf{0.0166} \\
Proposed & \textbf{0.0145} & \textbf{0.0336} & 0.0827 &  & \textbf{0.0037} & \textbf{0.0121} & 0.0209 &  & \textbf{0.0101} & \textbf{0.0207} & 0.0493 &  & \textbf{0.0042} & \textbf{0.0061} & 0.0219 \\ \midrule 
 PPP = 4\\
Manipop& - & - & - &   & 0.0145 & 0.0503 & 0.0528 & & - & - & - &  & 0.0241 & 0.1287 & 0.1103 \\
Halimi & 0.0452 & 0.0211 & 0.2444 &  & 0.0046 & \textbf{0.0030} & 0.0508 &  & 0.0229 & 0.0174 & 0.1636 &  & 0.0028 & \textbf{0.0019} & 0.0159 \\
Lindell & 0.0239 & 0.0725 & 0.0428 &  & 0.0787 & 0.5421 & 0.0881 &  & 0.0273 & 0.1003 & 0.0419 &  & 0.0735 & 0.4967 & 0.0822 \\
Peng & 0.0073 & 0.0161 & \textbf{0.0275} &  & 0.0043 & 0.0056 & \textbf{0.0112} &  & 0.0068 & 0.0132 & \textbf{0.0193} &  & 0.0041 & 0.0046 & \textbf{0.0108} \\
Proposed & \textbf{0.0052} & \textbf{0.0141} & 0.0326 &  & \textbf{0.0026} & 0.0057 & 0.0150 &  & \textbf{0.0050} & \textbf{0.0084} & 0.0253 &  & \textbf{0.0027} & \textbf{0.0019} & 0.0153 \\ \midrule 
 PPP = 16\\
Manipop& 0.0930 & 0.0643 &0.2171 &    & 0.0038 & 0.0039 & 0.0261 & & 0.1753 &  0.0631 & 0.3271 &  & 0.0063 & 0.0060 & 0.0545 \\
Halimi & 0.0097 & \textbf{0.0065} & 0.1009 &  & 0.0024 & \textbf{0.0010} & 0.0334 &  & 0.0040 & \textbf{0.0023} & 0.0390 &  & \textbf{0.0013} & \textbf{0.0008} & \textbf{0.0045} \\
Lindell & 0.0296 & 0.0669 & 0.0511 &  & 0.0280 & 0.0518 & 0.0455 &  & 0.0253 & 0.1005 & 0.0430 &  & 0.0212 & 0.0528 & 0.0312 \\
Peng & 0.0044 & 0.0067 & \textbf{0.0144} &  & 0.0031 & 0.0032 & \textbf{0.0114} &  & 0.0043 & 0.0051 & \textbf{0.0120} &  & 0.0027 & 0.0026 & 0.0065 \\
Proposed & \textbf{0.0029} & 0.0082 & 0.0226 &  & \textbf{0.0019} & 0.0014 & 0.0126 &  & \textbf{0.0028} & 0.0027 & 0.0174 &  & 0.0019 & \textbf{0.0008} & 0.0129 \\ \bottomrule

\end{tabular}
\label{tab:quantiative}
\end{table*}

\subsection{Results on simulated data}

In this experiment, we use the same simulated dataset and the evaluation metrics described in the previous subsection. 

\myhead{Comparison methods.} We compare the proposed model to existing reconstruction methods without additional sensor fusion. We consider a state-of-the-art statistical method called Manipop~\cite{tachella2019bayesian} and the underlying iterative Bayesian method in Algorithm~\ref{alg1} by Halimi et al.~\cite{halimi2021robust}. In Algorithm~\ref{alg1}, we consider the same filter size consistent to the case of $L=4$ in Table~\ref{tab:procedure}. We also compare to two state-of-the-art deep learning models: Lindell et al.~\cite{lindell2018singlephoton} and Peng et al.~\cite{peng2020photonefficient}. We use the publicly available pre-trained model for~\cite{lindell2018singlephoton} and as no pre-trained model is available for \cite{peng2020photonefficient}, we train this model using the authors' publicly available codes. We also report the result of the classical algorithm obtained by applying a matched filter to the Lidar data by the system IRF.

\myhead{Qualitative comparison.} Fig.~\ref{fig:depth1} shows the reconstructed depth maps on the Art scene. In the case of high PPP and SBR, all the methods reconstruct well except for Lindell et al.~\cite{lindell2018singlephoton} which loses some details. In the challenging data case, we first notice that Manipop is conservative and shows many zero pixels to indicate the absence of a target in them. Other algorithms detect more targets, with the proposed algorithm showing the best robustness to outliers. These results are confirmed in
Fig.~\ref{fig:pc1} showing the point cloud representation of the reconstruction results. When PPP=4 and SBR=1, the previous deep learning methods suffer from so-called flying pixel artifacts (also called bleeding effects)~\cite{tosi2021smdnets} around the surface boundaries, while less artifacts are observed in Halimi et al.~\cite{halimi2021robust} and in our method. When PPP=1 and SBR=0.25, Halimi et al.~\cite{halimi2021robust} yields many outliers, comapred to the proposed method.  These two methods estimate similar uncertainty maps, however, the proposed method indicates higher uncertainty in noisy regions (e.g., see region behind the cone for PPP=4, SBR=1).  
Consistent results are observed for the Reindeer scene in Fig.~\ref{fig:depth2} and Fig.~\ref{fig:pc2}.

\myhead{Quantitative comparison.} In Table~\ref{tab:quantiative}, we quantitatively evaluate our model in different levels of PPP and SBR. Peng et al.~\cite{peng2020photonefficient} overall outperforms other methods in terms of RMSE, but it yields high errors in DAE and SEE. One reason is due to oversmoothing artifacts around the boundaries of surfaces. On the other hand, our method yields the lowest errors when PPP=1 in terms of DAE and SEE. When PPP is 16, both Halimi~\cite{halimi2021robust} and the proposed method show an overall good performance in terms of DAE and SEE. Manipop shows good performance for clean data at PPP=16 and SBR=4, but its errors are  higher in challenging cases, because Manipop sets as zero non-target regions. Although the used metrics are not fair to Manipop, we report the errors for reference and we empty the numbers when they are not meaningful.

\def\fh{250pt}
\begin{figure}[ht]
\centering
\includegraphics[totalheight=\fh]{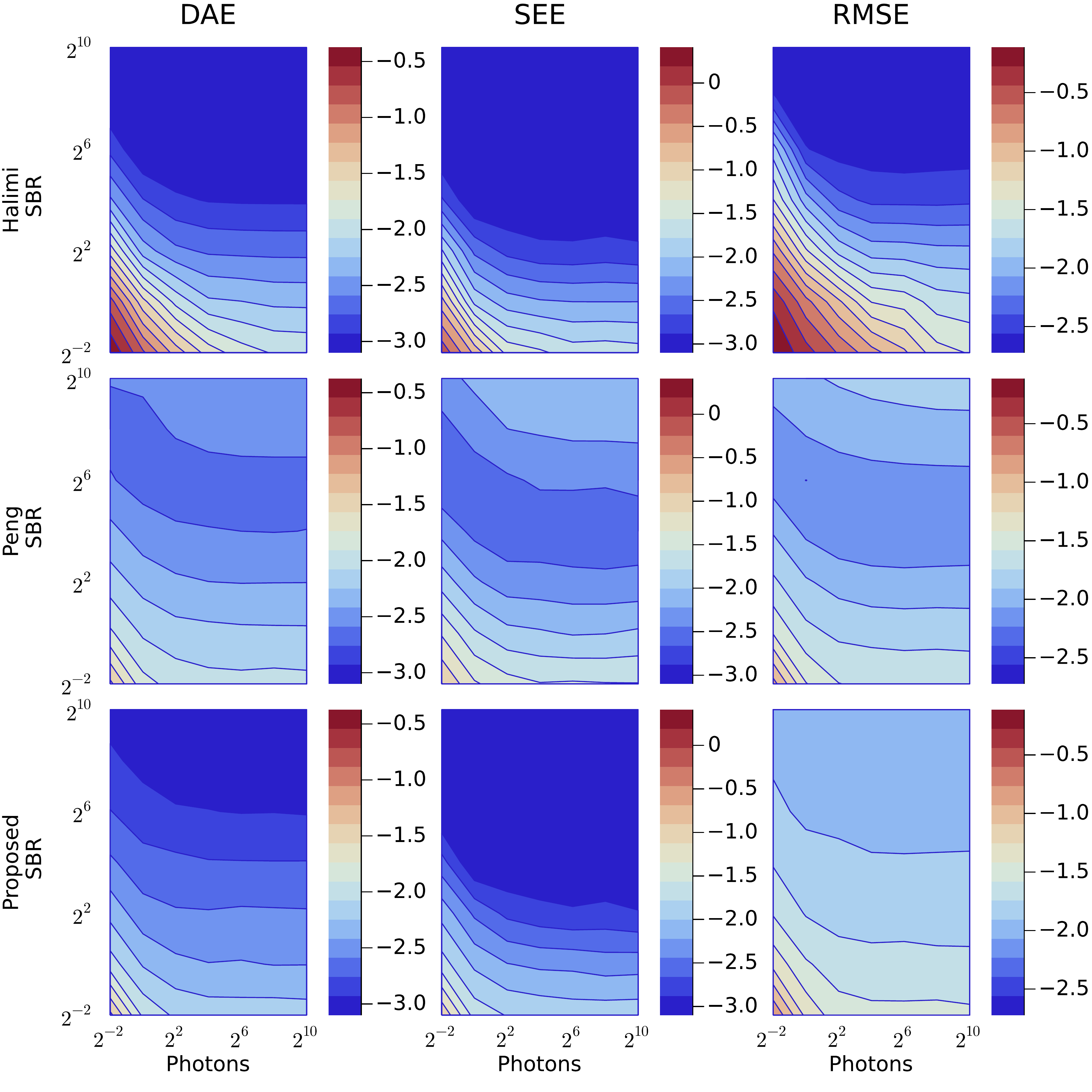}
\caption{Errors in terms of different levels of SBR and PPP on the Art scene by three methods: Halimi, Peng and the proposed (Top-to-Bottom). Three evaluation metrics of DAE, SEE and RMSE are used (Left-to-Right) and the error values are presented in a base-10 log scale.}
\label{fig:sbr_vs_ppp}
\end{figure}

As shown in Fig.~\ref{fig:sbr_vs_ppp}, we further conduct an extensive experiment in a wide range of PPP and SBR levels when comparing with Halimi et al~\cite{halimi2021robust} and Peng et al.~\cite{peng2020photonefficient}. The underlying Bayesian method~\cite{halimi2021robust} gives an excellent performance on the clean data, but its performance rapidly degrades in the low-photon and high noise cases. In such cases, Peng~\cite{peng2020photonefficient} and our method show comparable results, but the errors in Peng's method begin to increase when PPP is higher than 16 and SBR is higher than 512. Compared to Peng's result, the proposed method offers a more consistent performance even when PPP and SBR are high.

\myhead{Generalizability on different system IRFs.} Unlike previous work~\cite{lindell2018singlephoton,peng2020photonefficient}, the proposed method incorporates the system IRF, as explained in Section~\ref{sec:estimation}. Here, we test how robust our method is to changes affecting the system IRF.  We consider two types of baseline IRFs with 15 non-zero time bins: a symmetric IRF given by the Gaussian function and a realistic asymmetric IRF. On these baseline IRFs, we apply a Gaussian smoothing with different standard deviations $\sigma_{\mathrm{IRF}}$ and use the resulting IRF to  generate test data on the Art scene with PPP=4 and SBR=4. The first row of Fig.~\ref{fig:irf} shows the shapes of different IRFs where  a large value of $\sigma_{\mathrm{IRF}}$ increases the IRF's width. The second row shows the errors with respect to $\sigma_{\mathrm{IRF}}$ when considering the compared networks (without retraining with the modified IRFs). The performance of our method is shown to be less affected by the different IRFs in both symmetric and asymmetric cases~\cite{lindell2018singlephoton}, \cite{peng2020photonefficient}. This result highlights the robustness of our method to the mismodelling of the system IRF.

\def\fh{165pt}
\begin{figure}[th]
\centering
\includegraphics[totalheight=\fh]{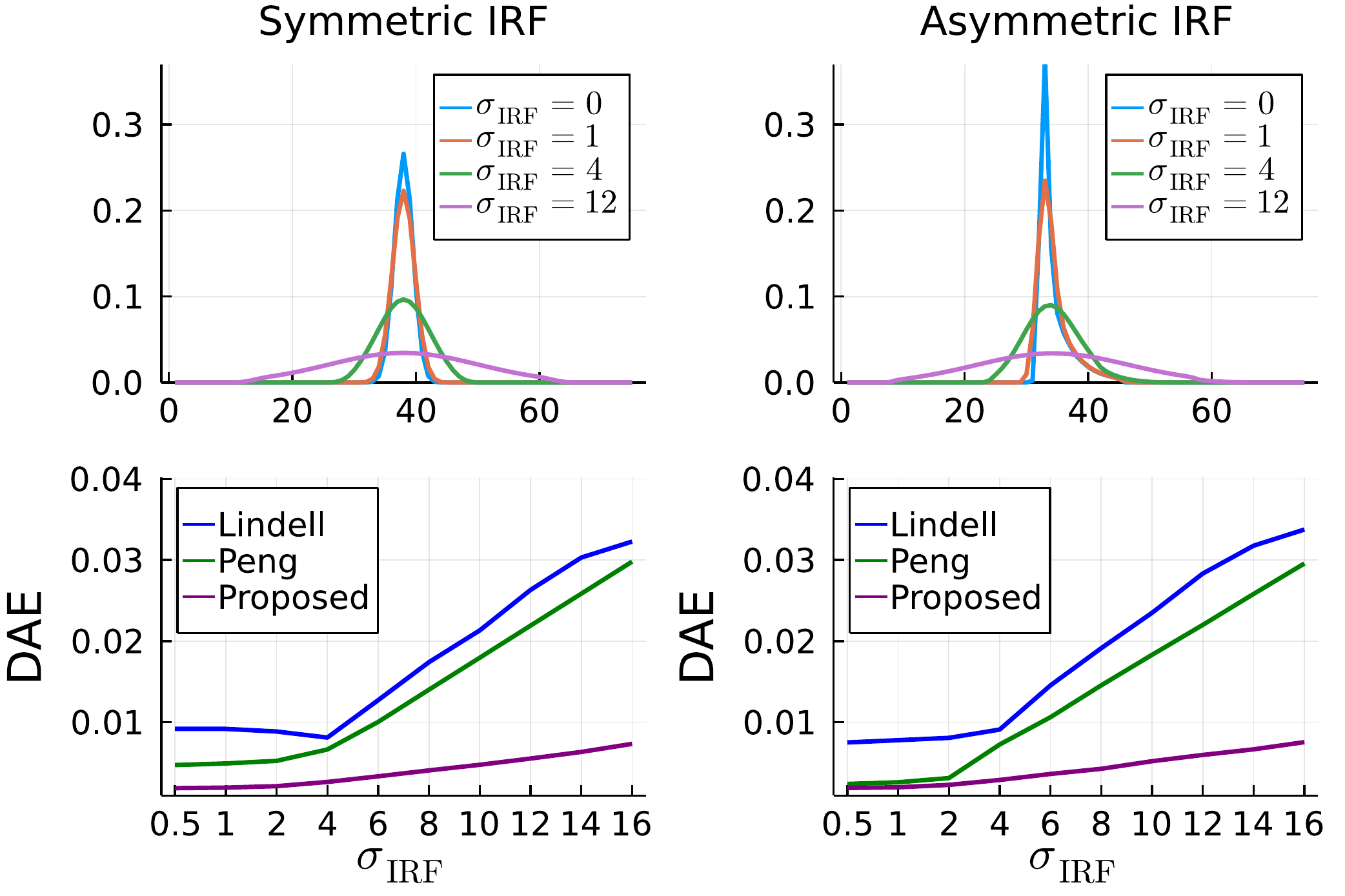}
\caption{Generalization test on different system IRFs. When generating test data, different system IRFs are considered by applying Gaussian smoothing with varying standard deviations $\sigma_{\mathrm{IRF}}$ on the two baseline IRFs: a Gaussian IRF (top-left) and a realistic asymmetric IRF (top-right). The bottom row shows the DAE by Lindell, Peng and the proposed method with varying $\sigma_{\mathrm{IRF}}$ from the Gaussian IRF (bottom-left) and the asymmetric IRF (bottom-right).}
\label{fig:irf}
\end{figure}

\begin{table}[ht]  \centering
 \caption{Running time on the Art scene with the resolution of 555$\times$695$\times$1024. The running time of the proposed method is presented into two parts: the estimation of initial multiscale depths and the inference by the network. For the deep learning methods, the training time is in the unit of hours.}
\begin{tabular}{l c r r r }
\toprule
\multicolumn{1}{c}{Method} &  Device & Running time (sec) & Train time & Parameters \\ \midrule
Classic & \multirow{3}{*}{CPU}  & 13.7 &  \\  
Manipop   &  & 2011.3 & \\
Halimi &   & 157.7 & \\ \midrule
Lindell  &  \multirow{2}{*}{GPU} & 427.6 & 24h & 1,728,996 \\
Peng  &   &  74.6 & 35h & 568,298 \\ \midrule
\multirow{2}{*}{Proposed} &  CPU &  317 + 4.7 = 322.1 & \multirow{2}{*}{9h} & \multirow{2}{*}{53,136} \\
&    GPU & 5.07 + 0.07 = 5.1 &  \\
\bottomrule
\end{tabular}
\label{tab:running_time}
\end{table}   

\def\fh{150pt}
\begin{figure}[ht]
\centering
\includegraphics[totalheight=\fh]{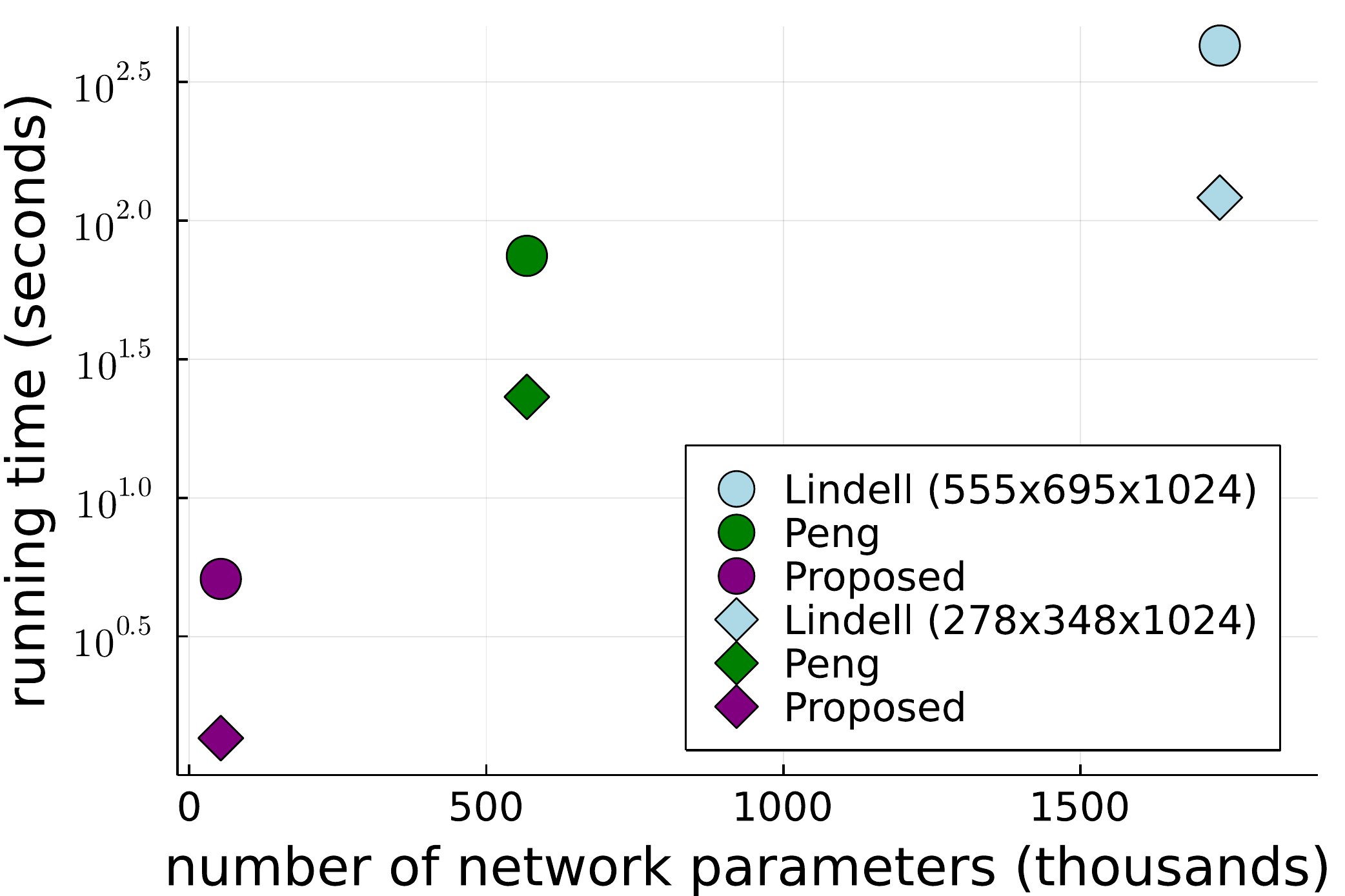}
\caption{The proposed model requires fewer parameters and lower running time in high dimensional Lidar data, compared to the state-of-the-art deep learning methods by Lindell~\cite{lindell2018singlephoton} and Peng~\cite{peng2020photonefficient}. The running time is tested on two Lidar data cube: one with a size of $555{\times}695{\times}1024$ (circles) and the other with $278{\times}348{\times}1024$ (diamonds).}
\label{fig:time_vs_params}
\end{figure}

\myhead{Efficiency of the network.} Table~\ref{tab:running_time} compares the number of parameters of the compared deep learning methods and their running times on the Art scene. The proposed method  shows the fastest running time when using a GPU device. The previous deep learning models \cite{lindell2018singlephoton}, \cite{peng2020photonefficient} could not take as input the full Lidar data due to the GPU memory limit, so they process small size patches and stitch the resulting depths together to obtain the final estimate. This is why they have a large running time with high dimensional data. Meanwhile, in the proposed method, most of the computational costs come from the estimation of initial multiscale depths, which takes 317 seconds on a CPU device and 5.07 seconds on a GPU device. The parameters of our method are an order of magnitude less than those of~\cite{lindell2018singlephoton},~\cite{peng2020photonefficient}, and hence the proposed method requires shorter training time. Fig.~\ref{fig:time_vs_params} provides a visual comparison of testing time in terms of the number of network parameters on two different sizes of data. 

It is worth mentioning that during training, Lindell and Peng use 13,800 patches of SPAD measurements with the size 32$\times$32$\times$1024 simulated from NYU v2 dataset~\cite{silberman2012indoor}, while the proposed method uses   7,860 patches of SPAD measurements with the size 256$\times$256$\times$1024 simulated from~\cite{hirschmuller2007evaluation}, \cite{butler2012naturalistic} where each patch is compressed into multiscale depths of size 256$\times$256$\times$12 to serve as an input to our  network.

\def\fh{43pt}
\def\seqa{checkerboard}
\def\seqb{elephant}
\def\seqc{lamp}
\def\seqd{roll}
\def\seqe{stairs_ball}
\begin{figure*}[ht]
\begin{center}
\centering
\begin{tabular}{c@{ }c@{ }c@{ }c@{ }c@{ }c@{ }c@{ }c@{ }c@{ }c@{ }c}
\includegraphics[totalheight=\fh]{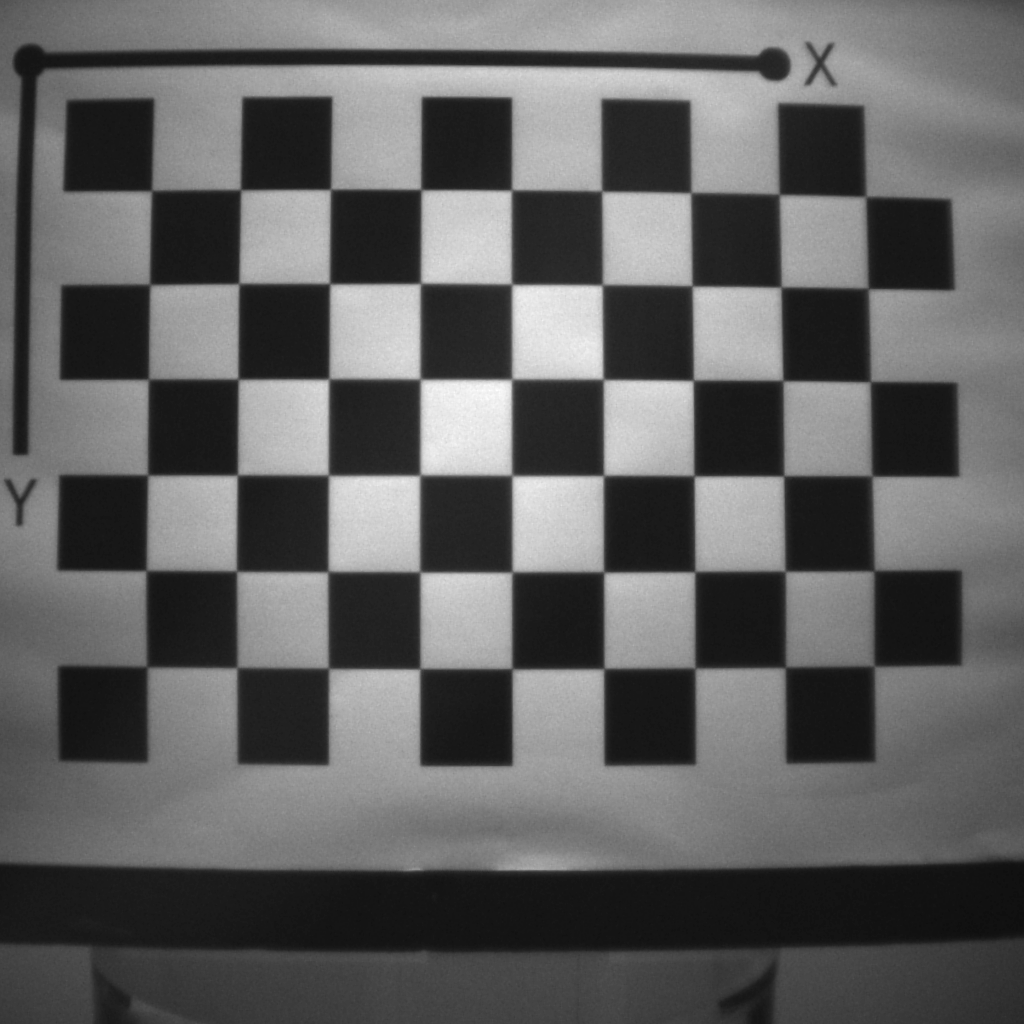} & 
\rotatebox[origin=l]{90}{\small\parbox{1.5cm}{\scriptsize$\,$PPP $= 3.6$\\SBR $\approx 0.8$}}&
\includegraphics[totalheight=\fh]{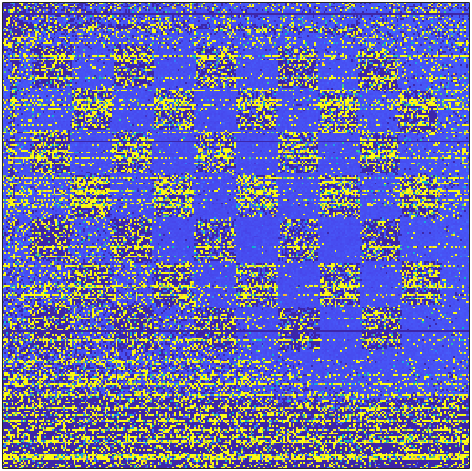} &
\includegraphics[totalheight=\fh]{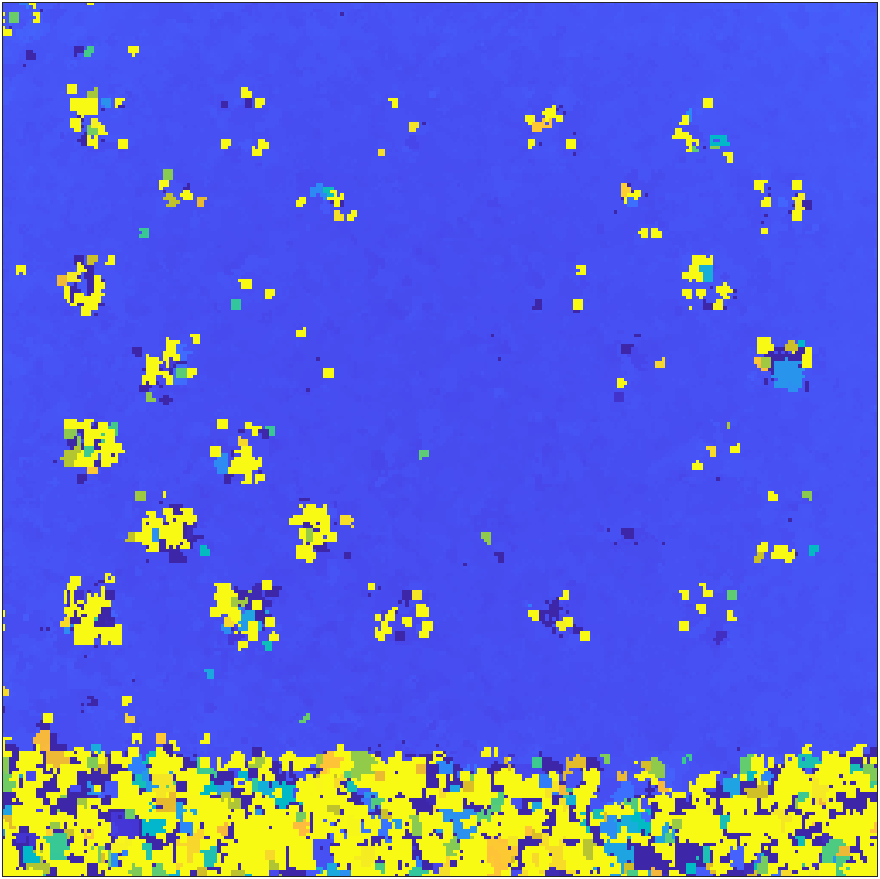} &
\includegraphics[totalheight=\fh]{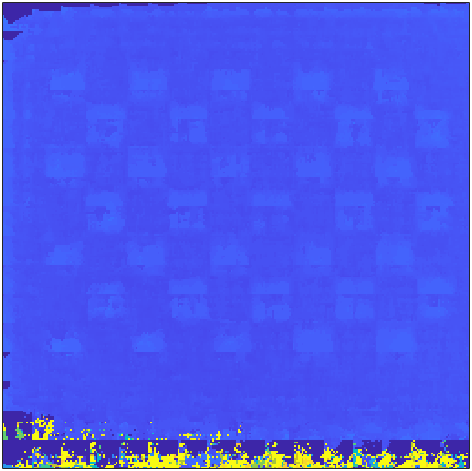} & 
\includegraphics[totalheight=\fh]{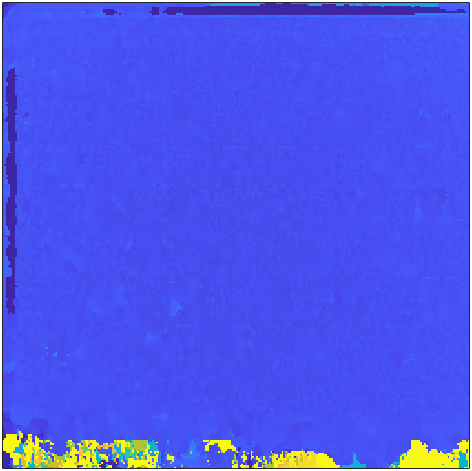} & 
\includegraphics[totalheight=\fh]{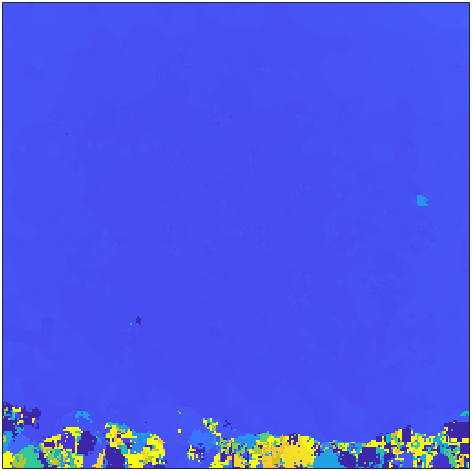} & \includegraphics[totalheight=\fh]{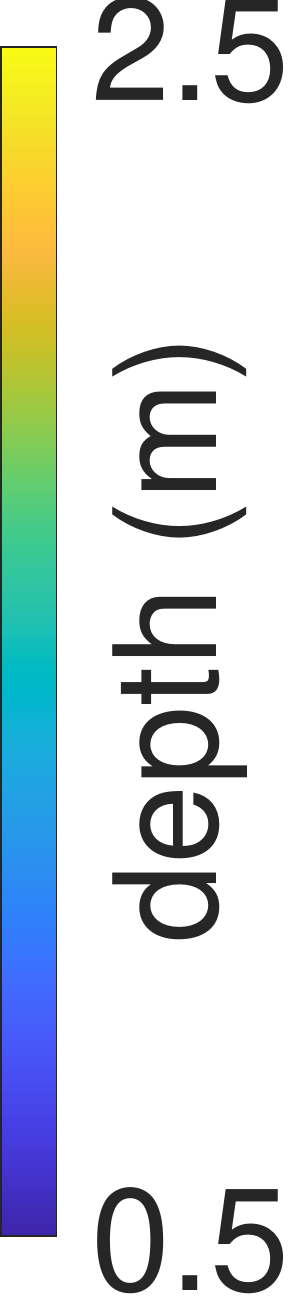} & \includegraphics[totalheight=\fh]{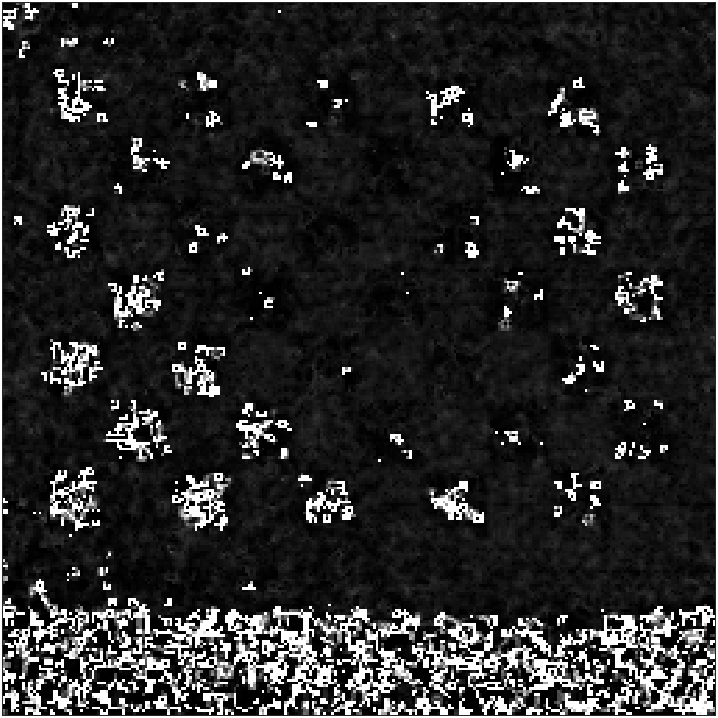} &
\includegraphics[totalheight=\fh]{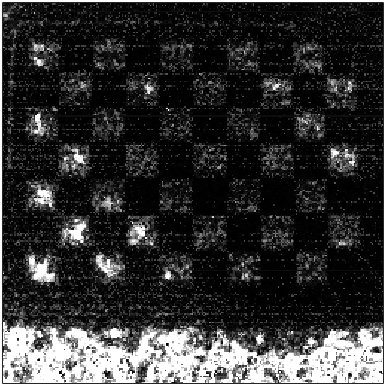} & \includegraphics[totalheight=\fh]{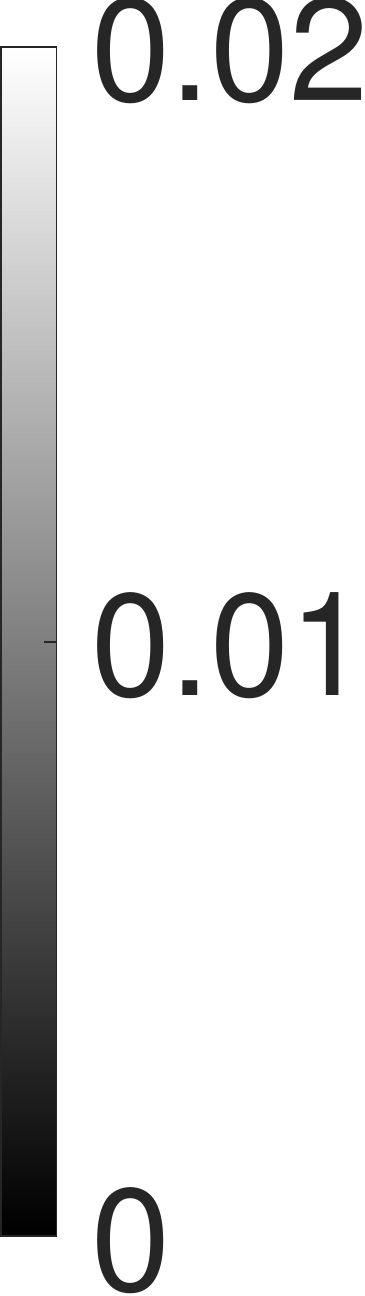}
\\

\includegraphics[totalheight=\fh]{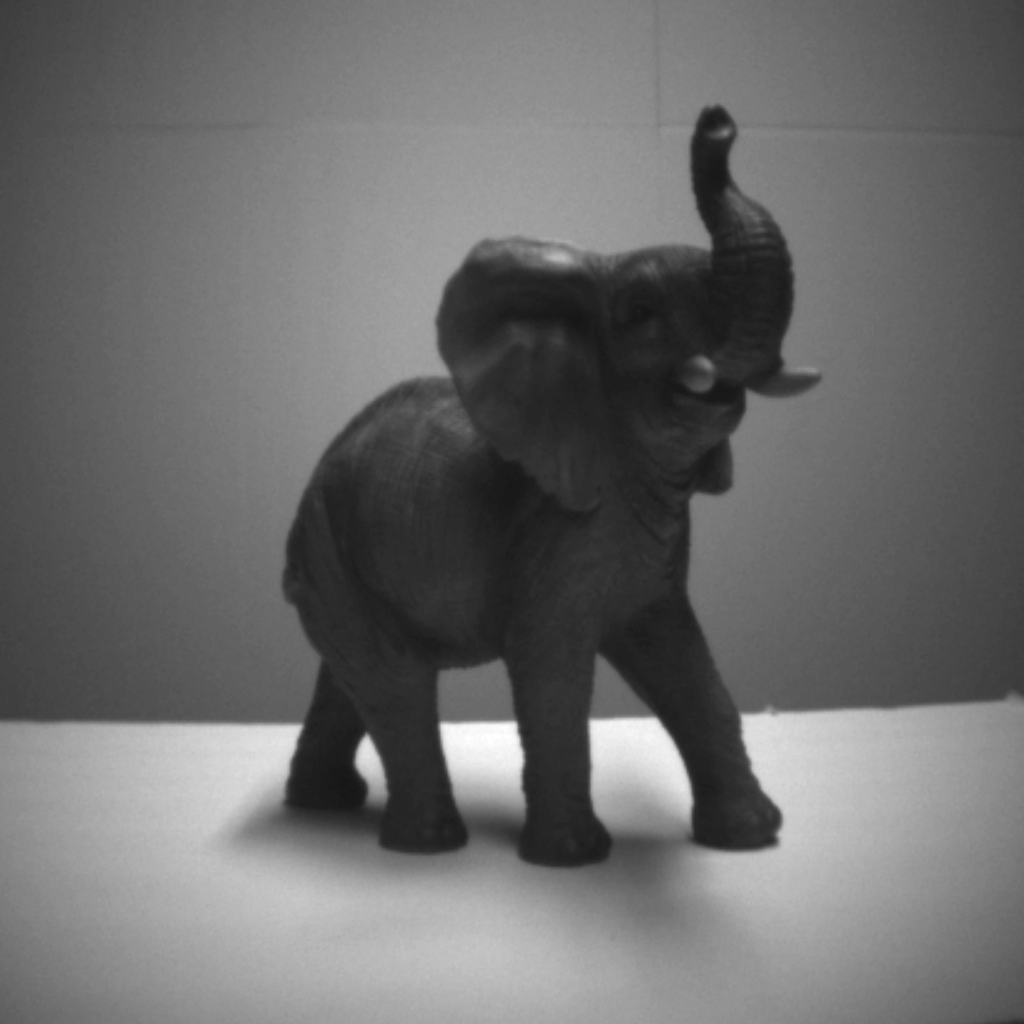} & 
\rotatebox[origin=l]{90}{\small\parbox{1.5cm}{\scriptsize$\,$PPP $= 2.5$\\SBR $\approx 0.35$}}&
\includegraphics[totalheight=\fh]{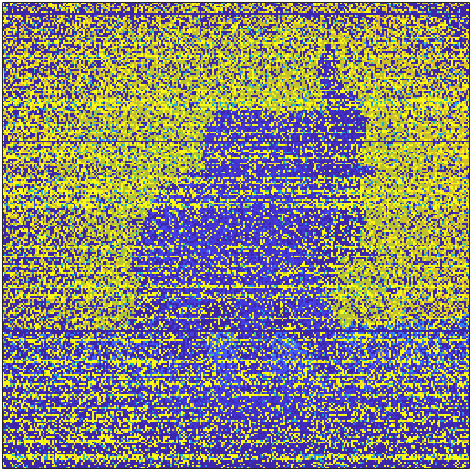} &
\includegraphics[totalheight=\fh]{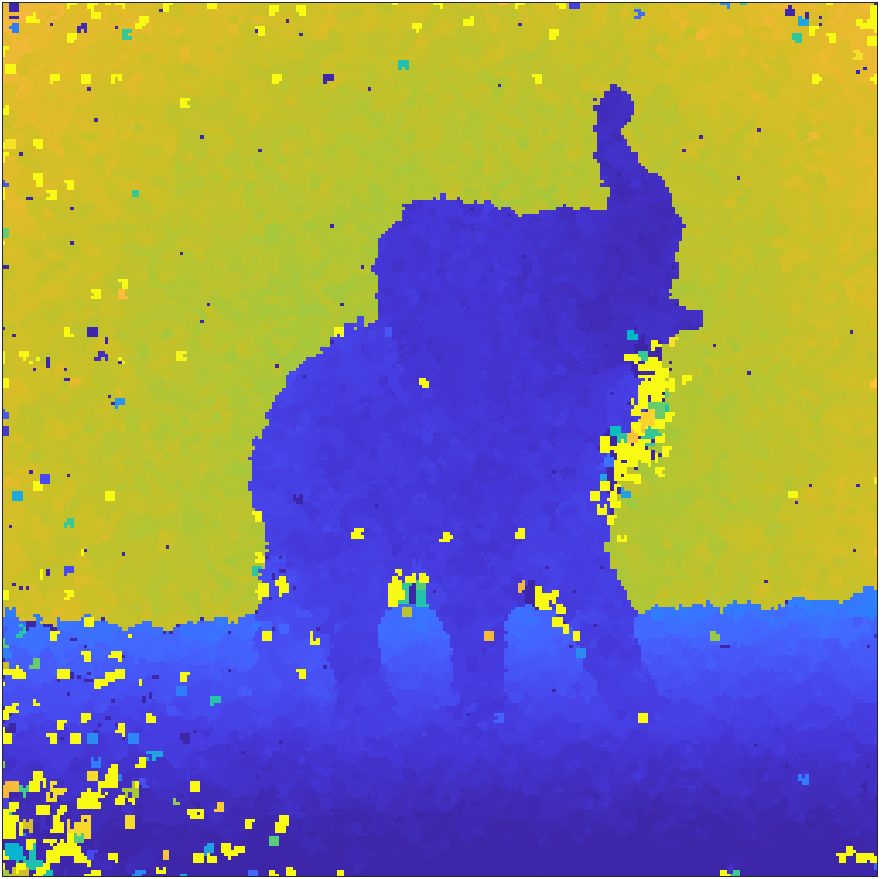} &
\includegraphics[totalheight=\fh]{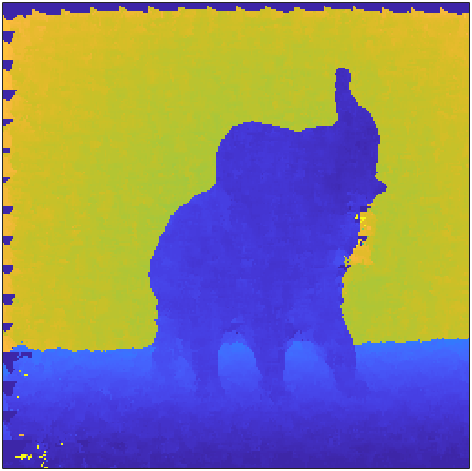} & 
\includegraphics[totalheight=\fh]{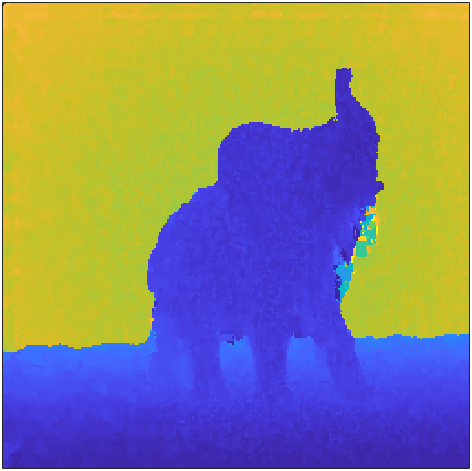} & 
\includegraphics[totalheight=\fh]{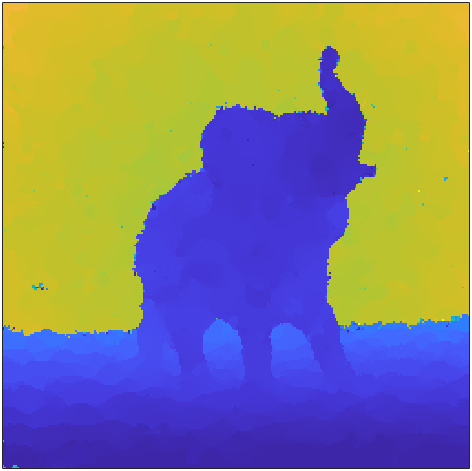} & \includegraphics[totalheight=\fh]{exp21/colorbar25} & \includegraphics[totalheight=\fh]{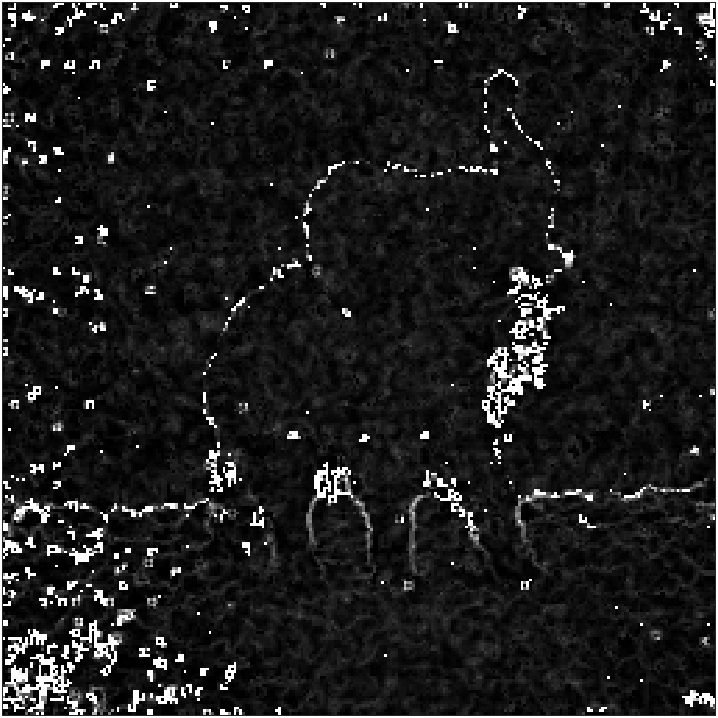} &
\includegraphics[totalheight=\fh]{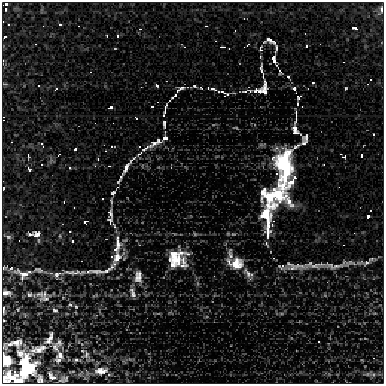} & \includegraphics[totalheight=\fh]{exp21/colorbar_uncertainty} \\

\includegraphics[totalheight=\fh]{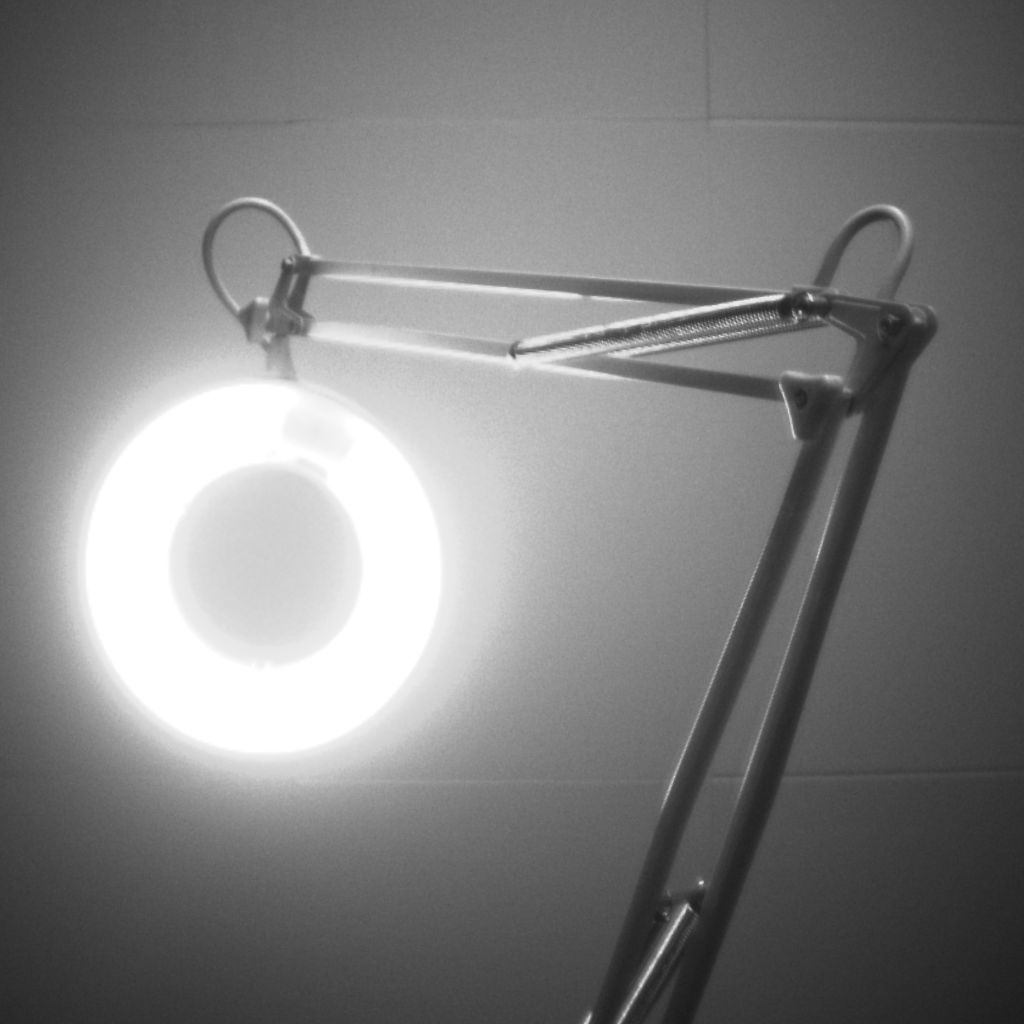} & 
\rotatebox[origin=l]{90}{\small\parbox{1.5cm}{\scriptsize$\,$PPP $= 3.0$\\SBR $\approx 0.3$}}&
\includegraphics[totalheight=\fh]{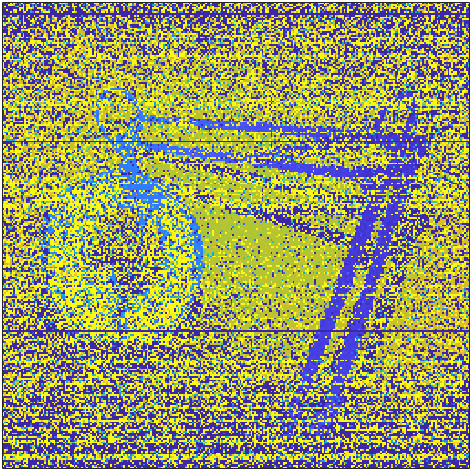} &
\includegraphics[totalheight=\fh]{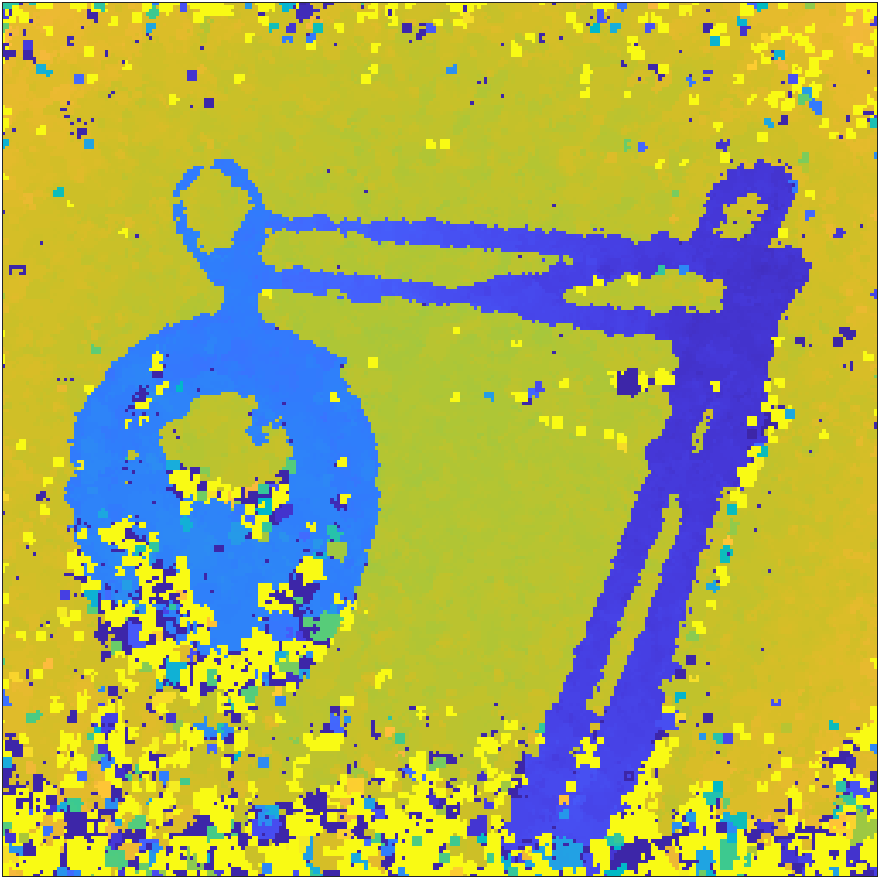} &
\includegraphics[totalheight=\fh]{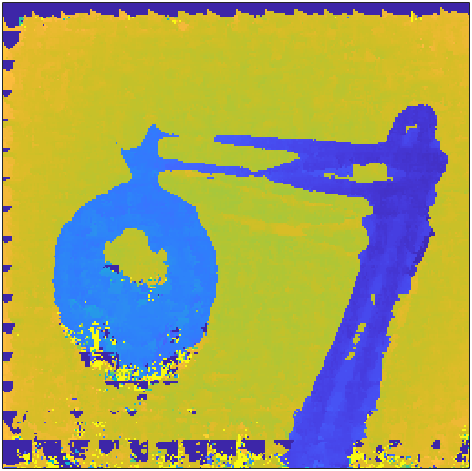} & 
\includegraphics[totalheight=\fh]{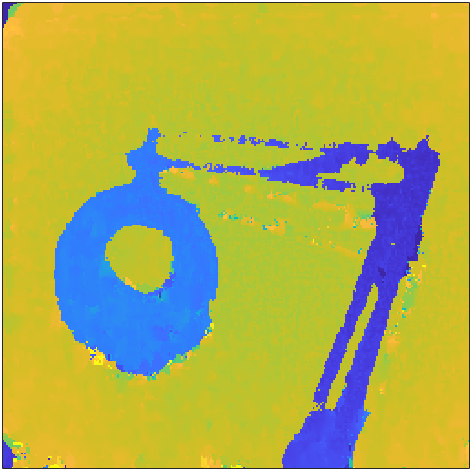} & 
\includegraphics[totalheight=\fh]{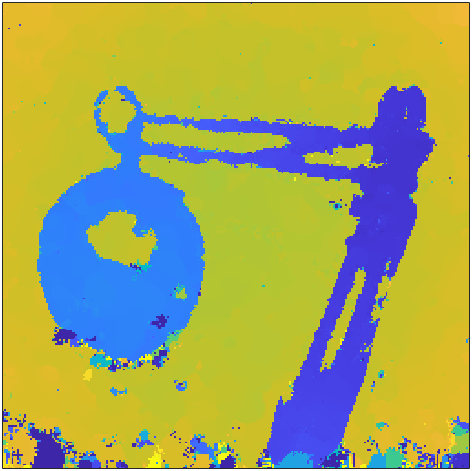} & \includegraphics[totalheight=\fh]{exp21/colorbar25} & \includegraphics[totalheight=\fh]{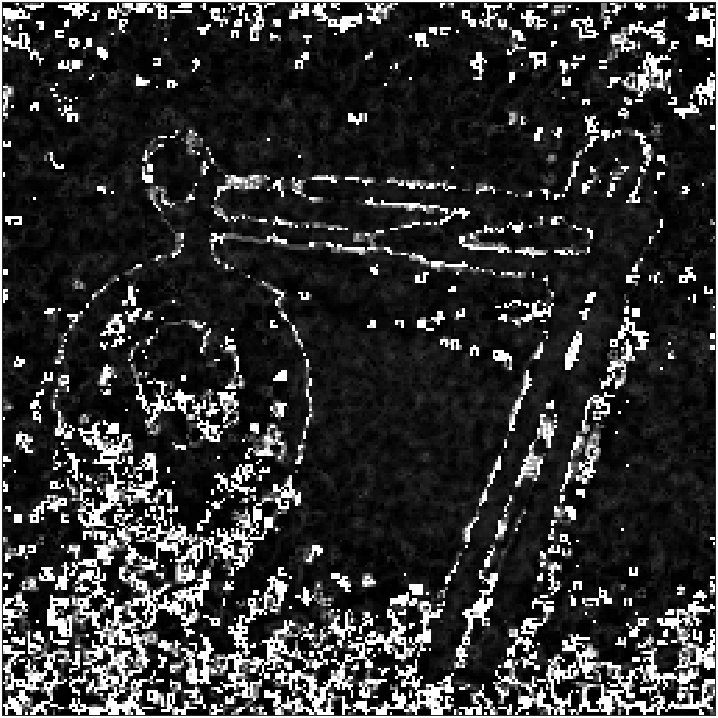} &
\includegraphics[totalheight=\fh]{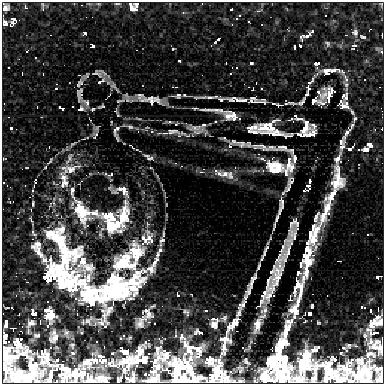} & \includegraphics[totalheight=\fh]{exp21/colorbar_uncertainty} \\

\includegraphics[totalheight=\fh]{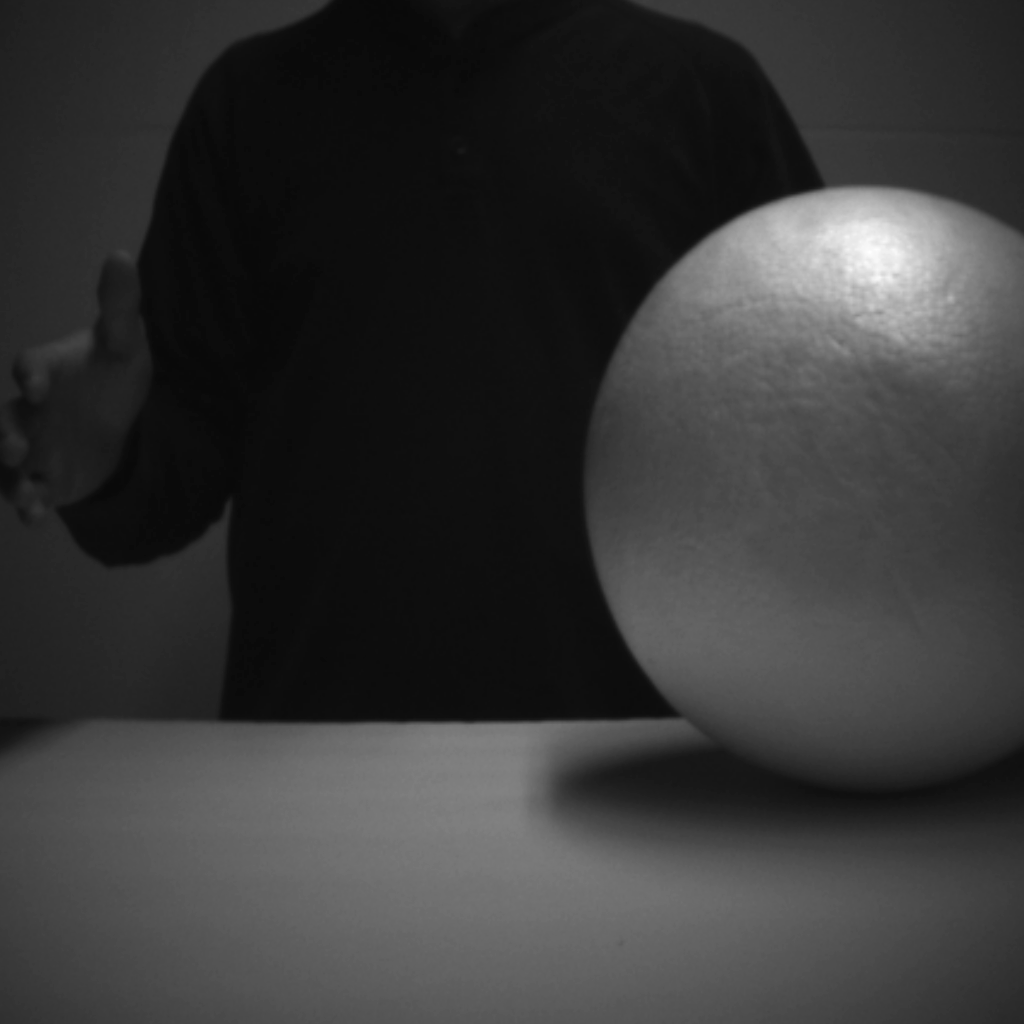} & 
\rotatebox[origin=l]{90}{\small\parbox{1.5cm}{\scriptsize$\,$PPP $= 2.8$\\SBR $\approx 0.56$}}&
\includegraphics[totalheight=\fh]{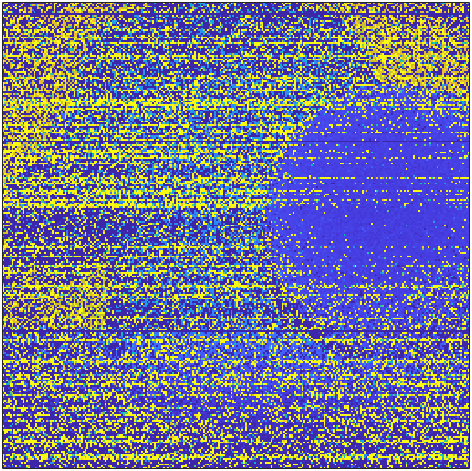} &
\includegraphics[totalheight=\fh]{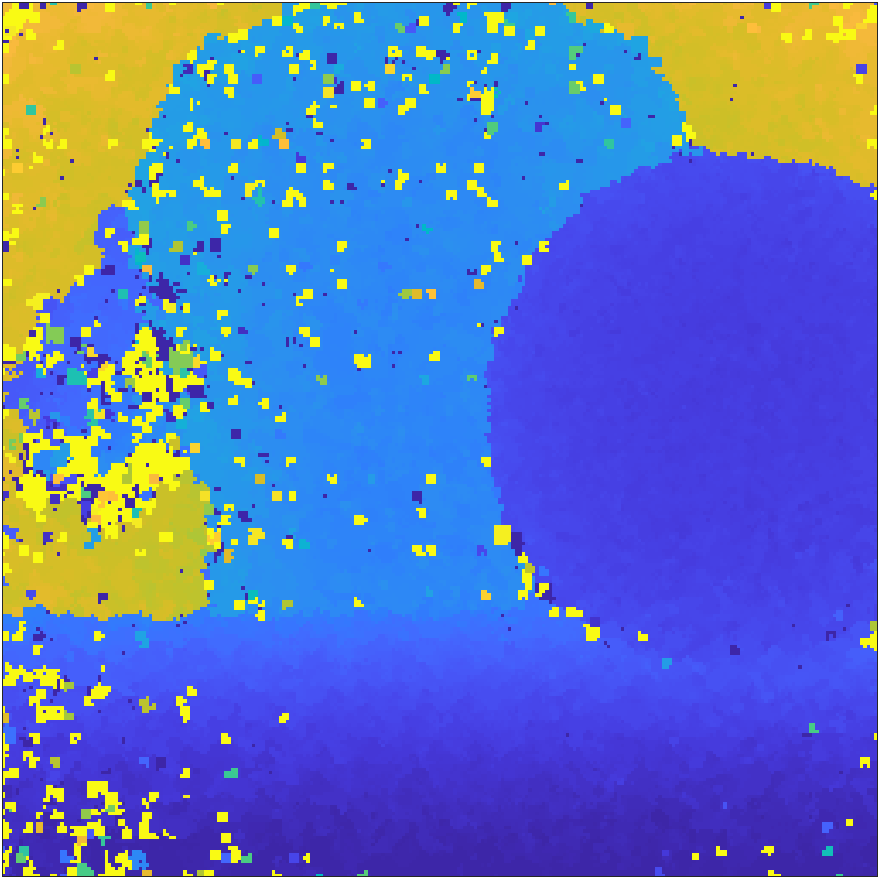} &
\includegraphics[totalheight=\fh]{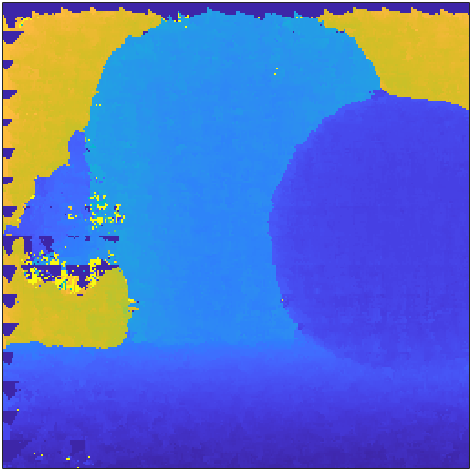} & 
\includegraphics[totalheight=\fh]{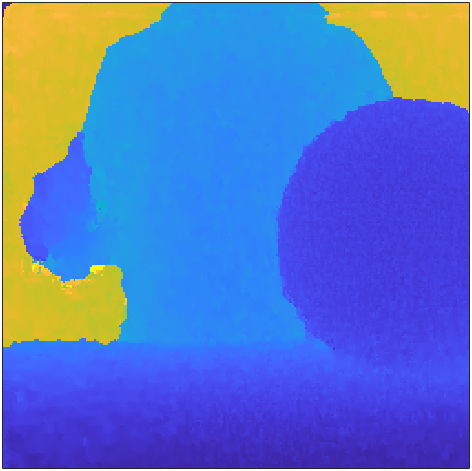} & 
\includegraphics[totalheight=\fh]{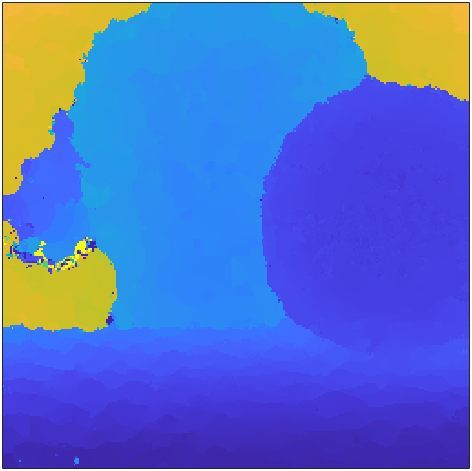} & \includegraphics[totalheight=\fh]{exp21/colorbar25} & \includegraphics[totalheight=\fh]{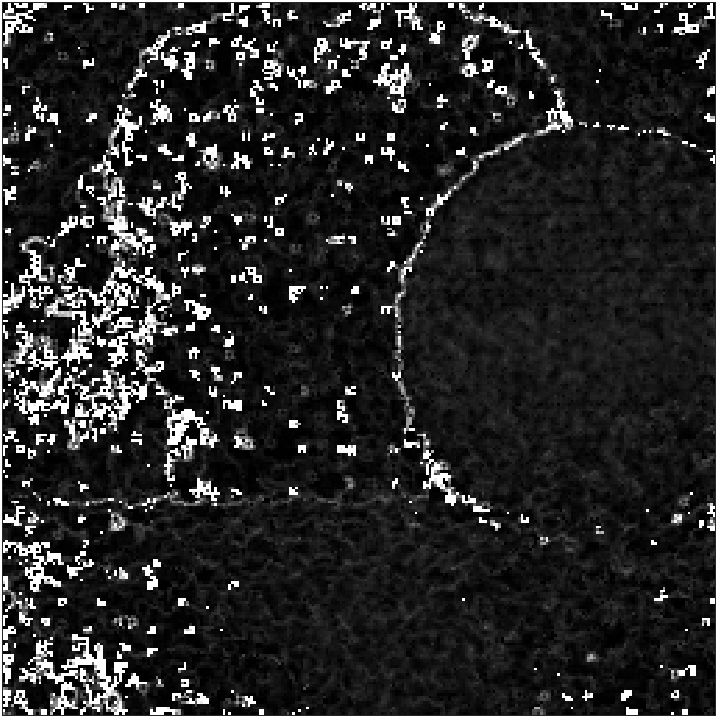} &
\includegraphics[totalheight=\fh]{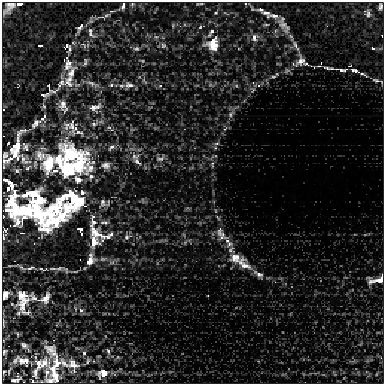} & \includegraphics[totalheight=\fh]{exp21/colorbar_uncertainty} \\

\includegraphics[totalheight=\fh]{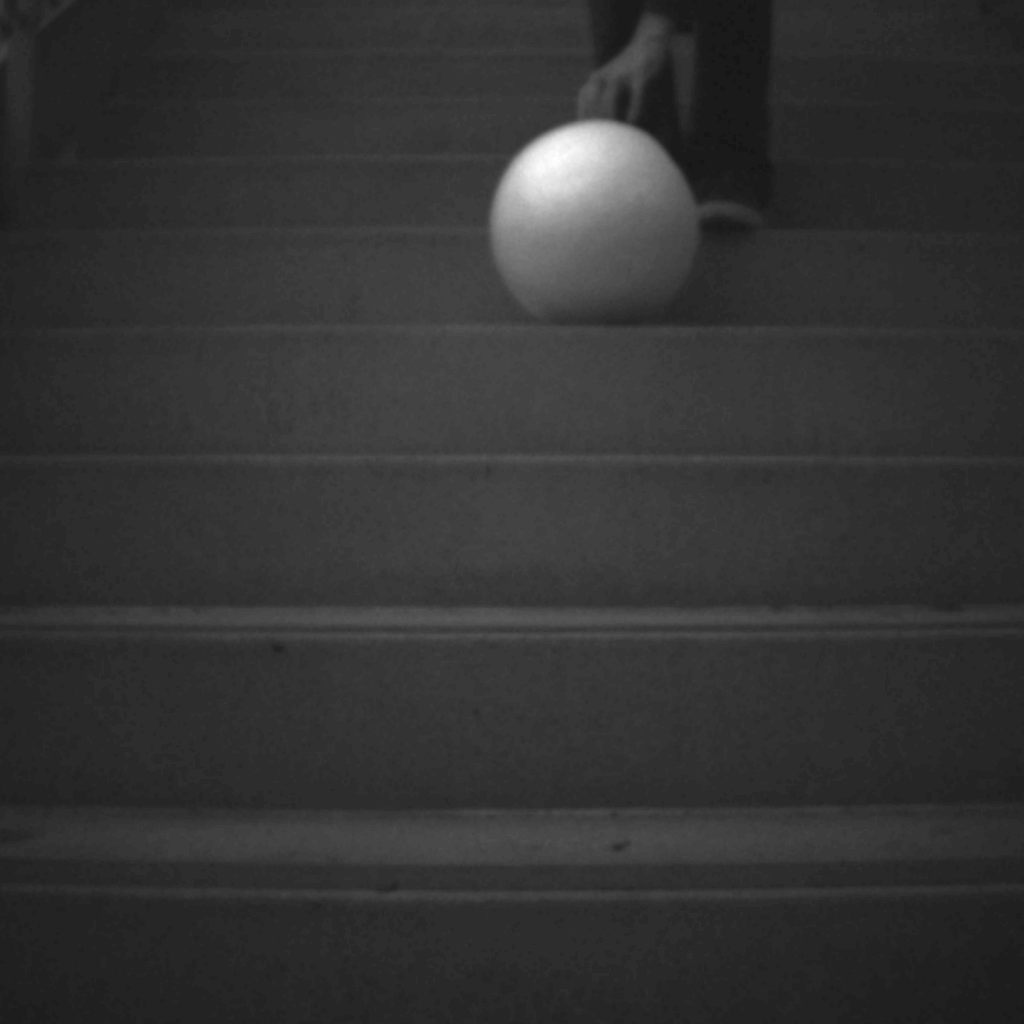} & 
\rotatebox[origin=l]{90}{\small\parbox{1.5cm}{\scriptsize$\,$PPP $= 11.1$\\SBR $< 0.07$}}&
\includegraphics[totalheight=\fh]{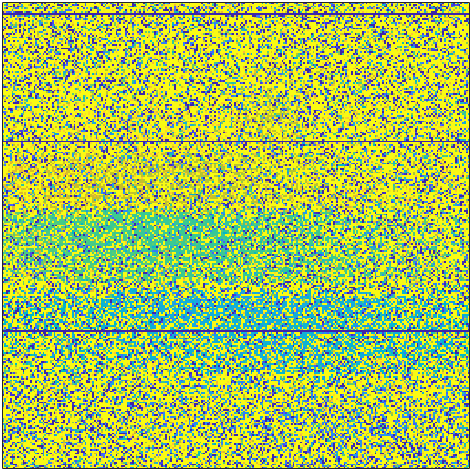} &
\includegraphics[totalheight=\fh]{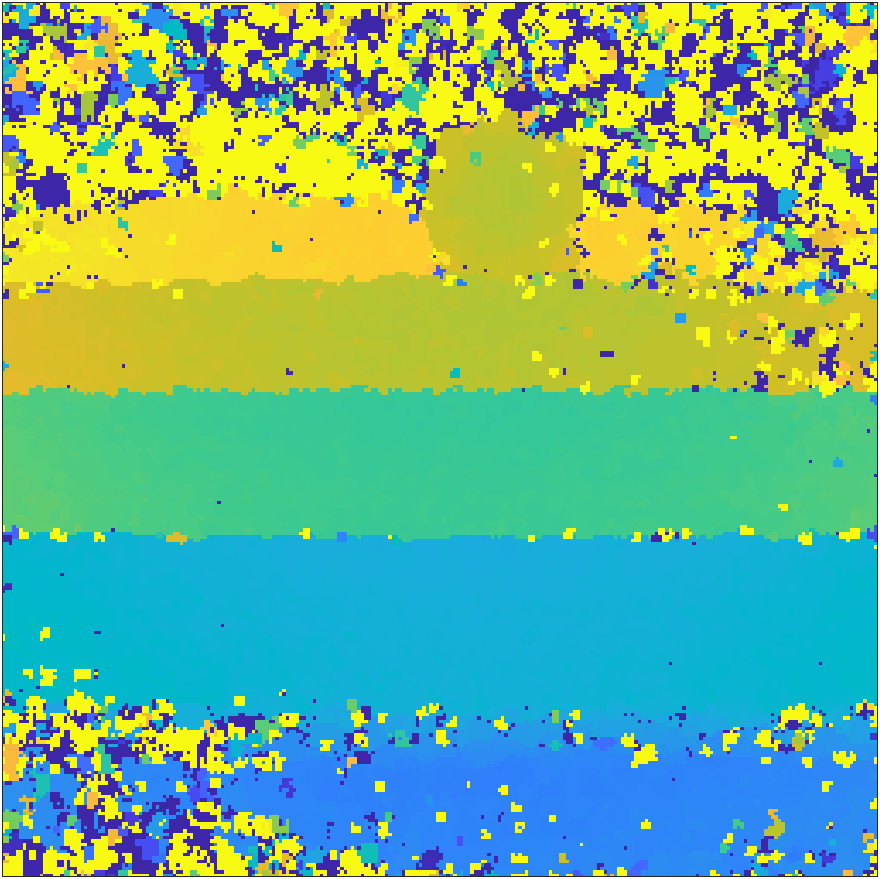} &
\includegraphics[totalheight=\fh]{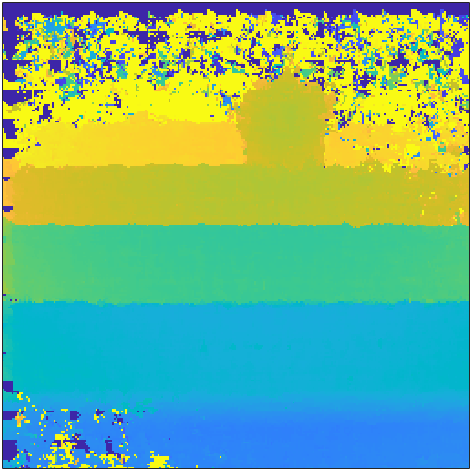} & 
\includegraphics[totalheight=\fh]{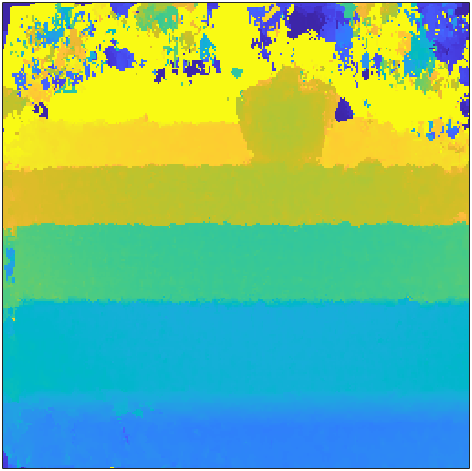} & 
\includegraphics[totalheight=\fh]{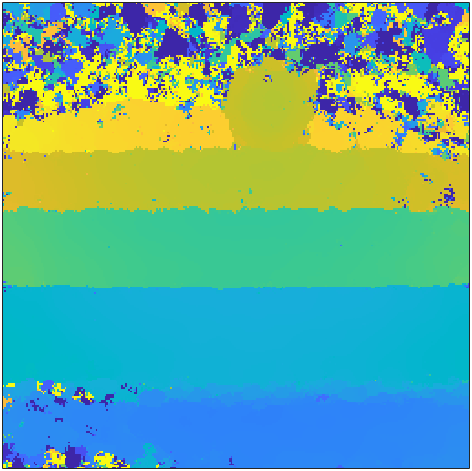} & \includegraphics[totalheight=\fh]{exp21/colorbar25}&\includegraphics[totalheight=\fh]{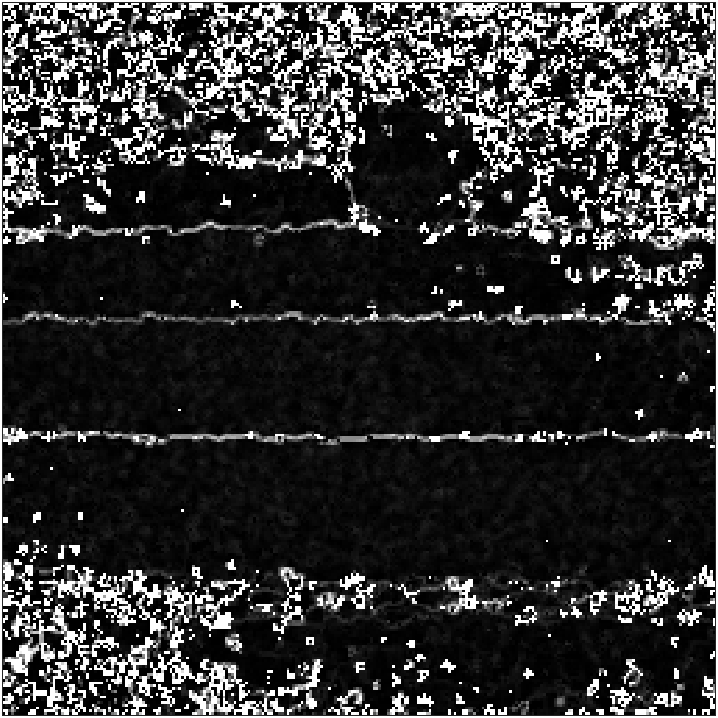}
& \includegraphics[totalheight=\fh]{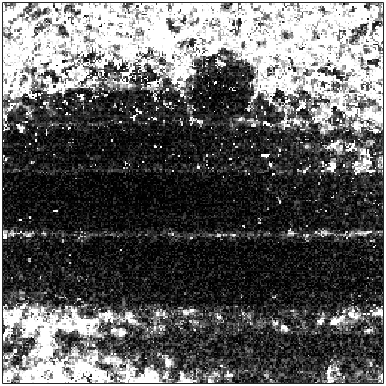} & \includegraphics[totalheight=\fh]{exp21/colorbar_uncertainty} \\
\multirow{2}{*}{Intensity} &&  \multirow{2}{*}{Classic} & \multirow{2}{*}{Halimi} & \multirow{2}{*}{Lindell} & \multirow{2}{*}{Peng} & \multirow{2}{*}{Proposed} & & Halimi & Proposed &  \\
&&&&&&&& \multicolumn{2}{c}{Uncertainty}  &
\end{tabular}
\caption{Reconstructed depth maps on the real dataset. The first column shows a reference intensity image and the last two columns show the uncertainty maps estimated by Halimi and the proposed method.}
\label{fig:indoor}
\end{center}
\end{figure*}
\def\fhpc{44pt}
\def\ftype{pc}
\begin{figure}[ht]
\centering
\begin{tabular}{c@{ }c@{ }c@{ }c@{ }c}
\includegraphics[totalheight=\fhpc]{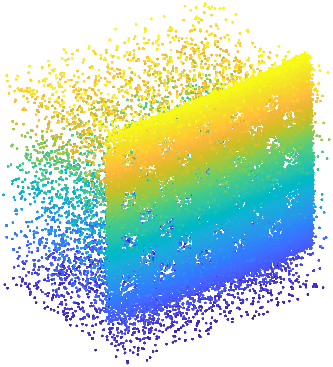} &
\includegraphics[totalheight=45pt]{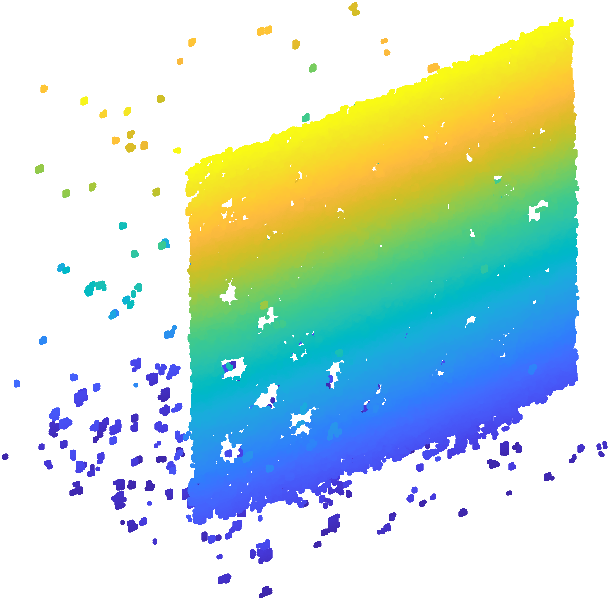} &
\includegraphics[totalheight=44pt]{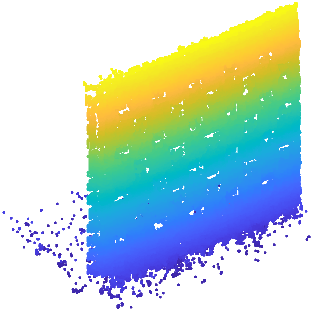} & 
\includegraphics[totalheight=44pt]{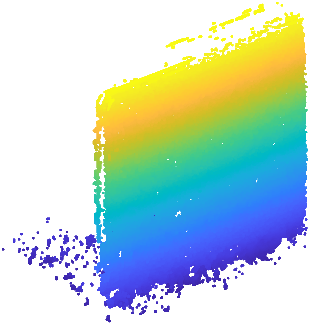} & 
\includegraphics[totalheight=44pt]{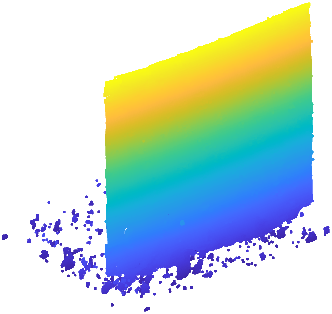} \\

\includegraphics[totalheight=\fhpc]{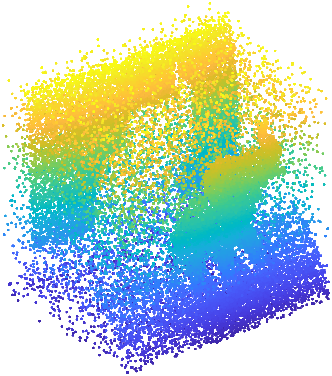} &
\includegraphics[totalheight=\fhpc]{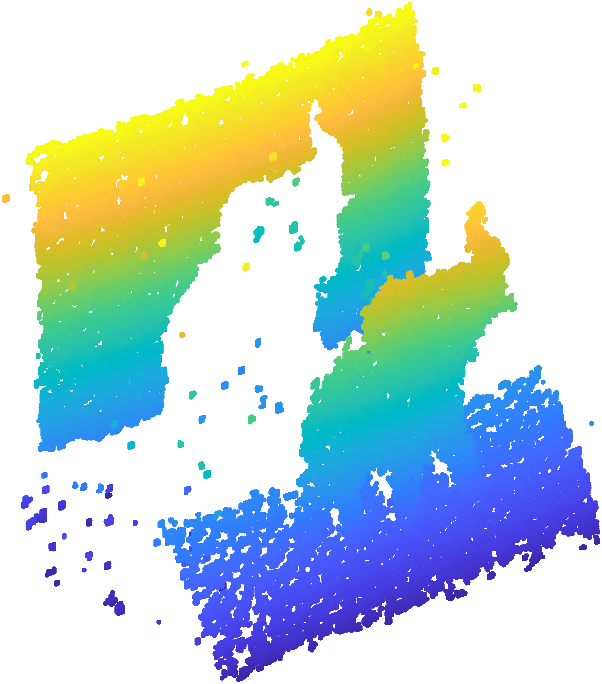} &
\includegraphics[totalheight=\fhpc]{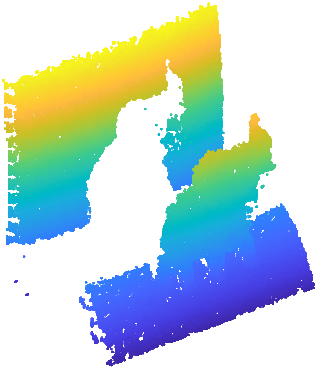} & 
\includegraphics[totalheight=\fhpc]{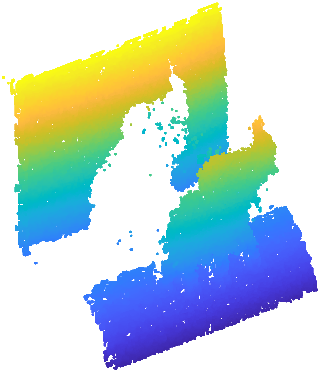} & 
\includegraphics[totalheight=\fhpc]{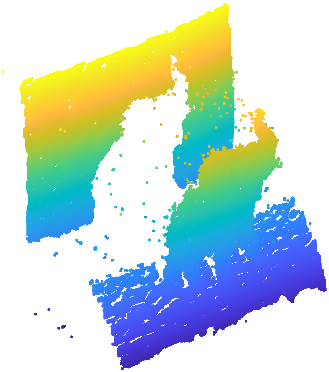} \\

\includegraphics[totalheight=\fhpc]{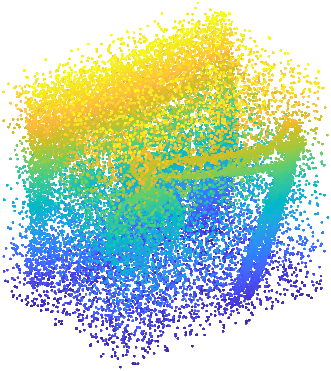} &
\includegraphics[totalheight=\fhpc]{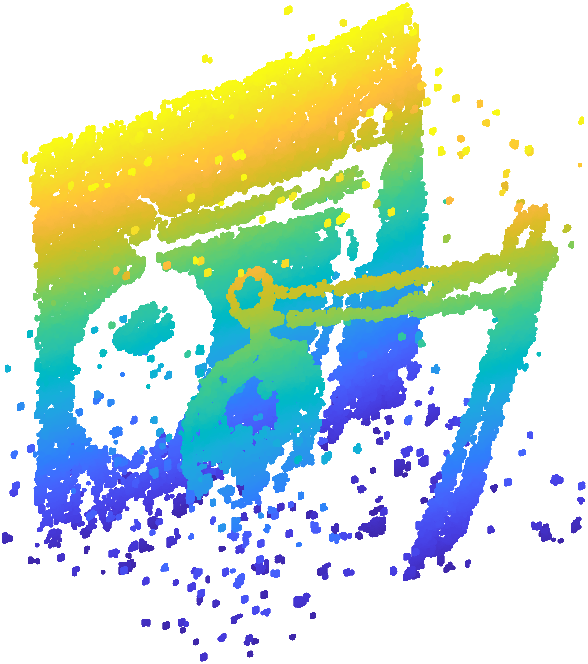} &
\includegraphics[totalheight=\fhpc]{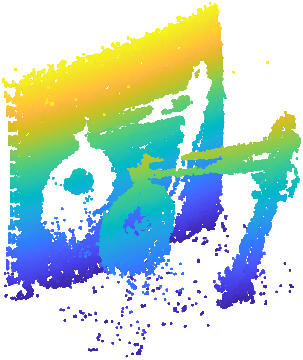} & 
\includegraphics[totalheight=\fhpc]{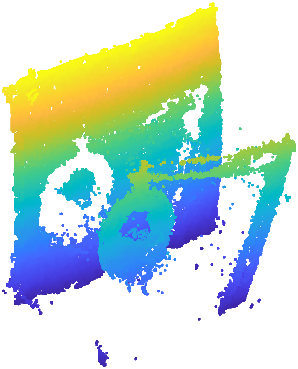} & 
\includegraphics[totalheight=\fhpc]{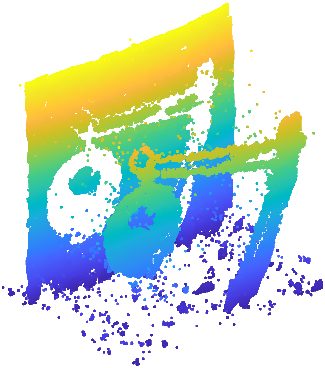} \\

\includegraphics[totalheight=\fhpc]{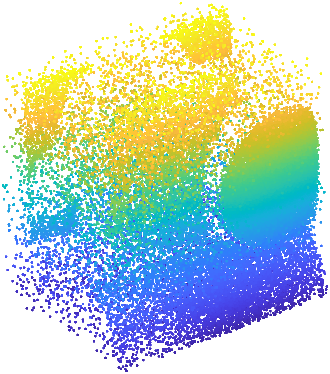} &
\includegraphics[totalheight=\fhpc]{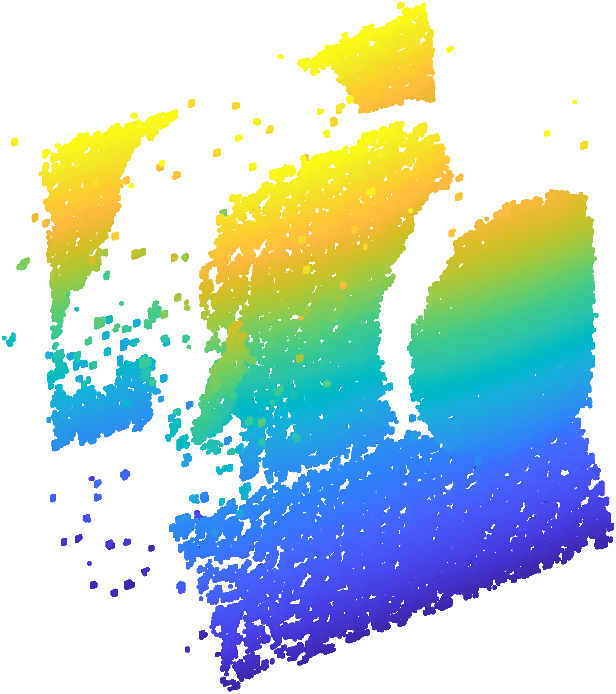} &
\includegraphics[totalheight=\fhpc]{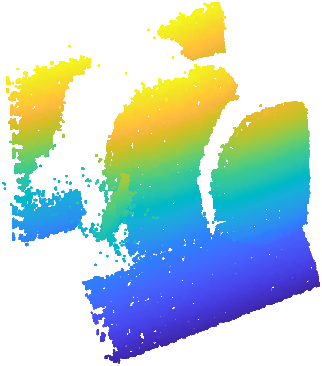} & 
\includegraphics[totalheight=\fhpc]{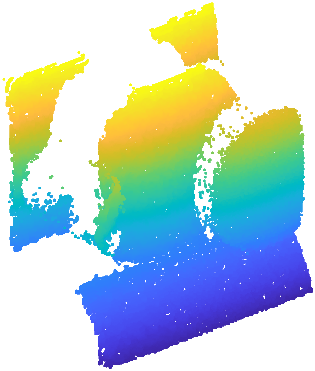} & 
\includegraphics[totalheight=\fhpc]{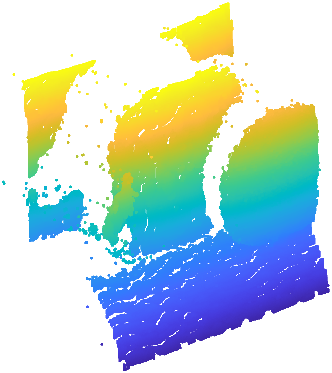} \\

\includegraphics[totalheight=\fhpc]{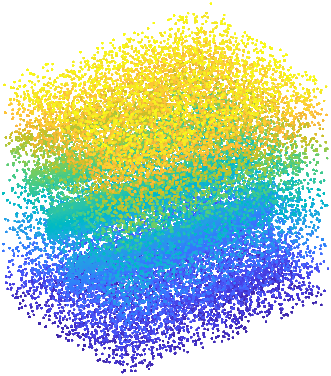} &
\includegraphics[totalheight=\fhpc]{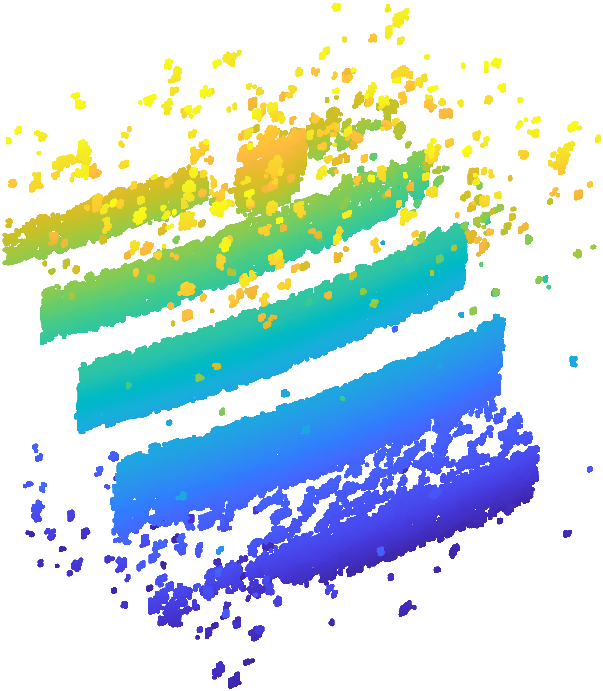} &
\includegraphics[totalheight=\fhpc]{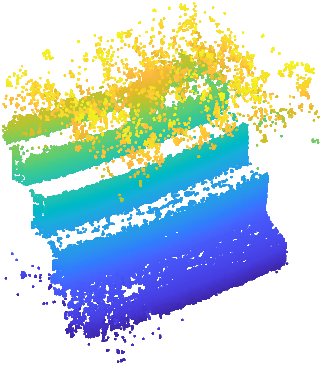} & 
\includegraphics[totalheight=\fhpc]{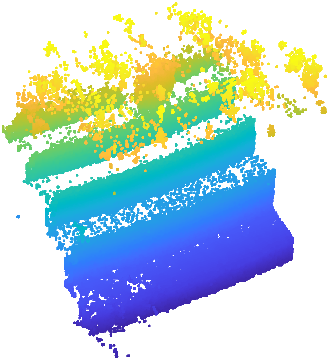} & 
\includegraphics[totalheight=\fhpc]{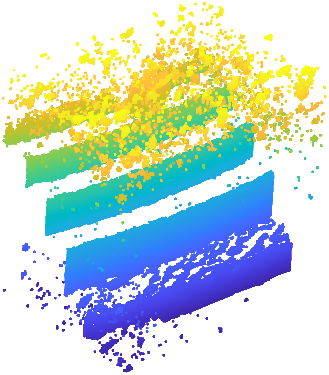} \\
Classic & Halimi & Lindell & Peng & Proposed 
\end{tabular}
\caption{Point cloud representation of reconstruction results on the real dataset.}
\label{fig:indoor_pc}
\end{figure}

\subsection{Results on real data} \label{sec:real}

We evaluate the proposed method on a real dataset provided in~\cite{lindell2018singlephoton} which captures real scenes under challenging scenarios. The Lidar data cubes have the resolution of 256$\times$256$\times$1536 and the first column of Fig.~\ref{fig:indoor} shows the reference intensity images of 4 indoor scenes (1st to 4th row) and 1 outdoor scene (the last row). In the figure, we report the PPP and SBR levels which are approximately estimated. Due to the high noise on the real data, Manipop does not yield meaningful surfaces, so we omit its results. As shown in the checkerboard scene (1st row), Peng~\cite{peng2020photonefficient} and our method yield flat depth maps on the checkerboard, while other methods observe inaccurate depth maps affected by the textures of the checkerboard. Compared to Peng's result, the proposed method gives a more flat depth map within the checkerboard and has less artifacts around the top and left borders. In the elephant scene (2nd row), compared to other methods, the proposed method reconstructs better the boundary of the elephant. In the lamp scene (3rd row), Halimi~\cite{halimi2021robust} reconstructs the structures well, but it suffers from outliers, while deep learning methods overall obtain less noisy results. Both Lindell~\cite{lindell2018singlephoton} and Peng~\cite{peng2020photonefficient} lose some details on the top of the lamp whereas the proposed method obtains better reconstruction. In the 4th row, Peng's method fails to capture the hand in the middle left region, while our method still captures it. Fig.~\ref{fig:indoor_pc} shows the reconstruction results represented by point clouds. The last row shows the reconstruction on the stair scene which has a very low SBR level due to strong sunlights. The previous deep learning methods often result in bleeding artifacts between the steps on the stair, while Halimi~\cite{halimi2021robust} and our method show fewer such artifacts. Finally, Table~\ref{tab:comptime} compares the running time, where the fastest results are obtained by the proposed method confirming its efficiency.
\vspace{-1em}
\begin{table}[!ht] \caption{Comparison of running time (in seconds) averaged on five real scenes with the resolution of 256$\times$256$\times$1536.}
\centering \ra{0.7}
\begin{tabular}{ccccc}  \toprule
Classic&Halimi&Lindell&Peng&Proposed \\ \midrule
4.1 (CPU) & 24.3 (CPU) & 222.1 (GPU) & 44.5 (GPU) & 1.2 (GPU) \\
\bottomrule
\end{tabular}
\label{tab:comptime}
\end{table}

%
\section{Conclusion and discussion} \label{sec:conclusion}

In this paper, we have proposed a new deep learning model to reconstruct depth profiles from single-photon Lidar data, taking advantages of statistical models and data-driven approaches. We design our neural network by unrolling a previous iterative Bayesian method~\cite{halimi2021robust}, exploiting the domain knowledge on a single-photon Lidar system. This unrolling strategy improves the interpretability and efficiency of the proposed network in terms of the network size, and the training and testing times. The resulting network is also more robust to mismodeling effects due to differences between training and testing data, than classical architectures.
The numerical experiments show that the proposed model can reconstruct high quality depth maps in challenging scenarios with less artifacts around the surface boundaries. Extending the model by accounting for the reflectivity maps as input is interesting and will be studied in the future.   




\renewcommand{\baselinestretch}{0.96}
\bibliographystyle{IEEEtran}
\bibliography{unroll}

\end{document}